\documentclass[prl,aps,twocolumn,superscriptaddress,floatfix,longbibliography,notitlepage,amssymb,amsmath]{revtex4-1}
\usepackage{array, multirow, graphicx}
\usepackage[dvipsnames]{xcolor}
\usepackage{mathtools}
\usepackage{comment}
\usepackage{epsfig}

\usepackage{hyperref}
\hypersetup{
    colorlinks=true,
    linkcolor=red,
    filecolor=magenta,      
    urlcolor=blue,
    citecolor = teal,
}

\newcommand{\beqn}{\begin{eqnarray}}
\newcommand{\eeqn}{\end{eqnarray}}
\newcommand{\eq}[1]{(\ref{#1})}

\newcommand{\cL}{{\cal L}}

\newcommand{\Z}{{\mathbb Z}}
\newcommand{\bs}{\boldsymbol}

\newcommand{\avr}[1]{{\left\langle #1 \right\rangle}}

\def\bbbone{{\mathchoice {\rm 1\mskip-4mu l} {\rm 1\mskip-4mu l} {\rm 1\mskip-4.5mu l} {\rm 1\mskip-5mu l}}}

\newcommand{\Tr}{{\mathrm{Tr}\,}}
\newcommand{\tr}{ {\rm Tr} \, }

\newcommand{\Arg}{{\mathrm{arg}\,}}

\newcommand{\us}[2]{\underset{#1}{#2}}

\newcommand{\lr}[1]{ \left( #1 \right) }

\newcommand{\lrm}[1]{ \left| #1 \right| }
\newcommand{\vev}[1]{ \left\langle #1 \right\rangle }
\newcommand{\hmu}{\hat{\mu}}
\newcommand{\hnu}{\hat{\nu}}

\newcommand{\EM}{\,{\mathrm{e.m.}}}
\newcommand{\GeV}{\,{\mathrm{GeV}}}

\newcommand{\EW}{\mathrm{EW}}

\begin{document}

\title{Phase structure of electroweak vacuum in a strong magnetic field: the lattice results}

\author{M. N. Chernodub}
\affiliation{Institut Denis Poisson UMR 7013, Universit\'e de Tours, 37200 Tours, France}
\author{V. A. Goy}
\affiliation{Pacific Quantum Center, Far Eastern Federal University, 690950 Vladivostok, Russia}
\author{A. V. Molochkov}
\affiliation{Pacific Quantum Center, Far Eastern Federal University, 690950 Vladivostok, Russia}

\begin{abstract}
Using first-principle lattice simulations, we demonstrate that in the background of a strong magnetic field (around $10^{20}$\,T), the electroweak sector of the vacuum experiences two consecutive crossover transitions associated with dramatic changes in the zero-temperature dynamics of the vector $W$ bosons and the scalar Higgs particles, respectively. Above the first crossover, we observe the appearance of large, inhomogeneous structures consistent with a classical picture of the formation of $W$ and $Z$ condensates pierced by vortices. The presence of the $W$ and $Z$ condensates supports the emergence of the exotic superconducting and superfluid properties induced by a strong magnetic field in the vacuum. We find evidence that the vortices form a disordered solid or a liquid rather than a crystal. The second transition restores the electroweak symmetry. Such conditions can be realized in the near-horizon region of the magnetized black holes.
\end{abstract}

\date{\today}

\bibliographystyle{apsrev4-1}

\maketitle

\paragraph*{\bf Introduction.}   
A powerful magnetic field background can modify the physical properties of the vacuum. For electromagnetic interactions described by Quantum Electrodynamics, the relevant intensity of the magnetic field is set by the Schwinger limit, $B^{\mathrm{QED}} = m_e^2/e \simeq 4 \times 10^9\,\mathrm{T}$~\footnote{We use the units $\hbar = c = 1$.} determined by the electron mass $m_e$~\cite{Schwinger:1951nm}. At this strength -- which is already bypassed by the fields near the surface of magnetars~\cite{Olausen2014mcgill} -- the vacuum acquires optical birefringence properties~\cite{Adler:1971wn} and can act as a ``magnetic lens'' which can distort and magnify images~\cite{Shaviv1999magnetic} similarly to the celebrated galaxy-scale gravitational lengths. 

Strong fundamental interactions, described by Quantum Chromodynamics, are affected by the magnetic field of the strength of the hadronic mass scale, $B^{\mathrm{QCD}} \sim m_p^2/e \sim 10^{16}\,\mathrm{T}$ where $m_p$ is the proton mass. Such fields generate the magnetic catalysis~\cite{Klevansky:1989vi, Klimenko:1991he, Shovkovy:2012zn}, which implies, in particular, a persistent enhancement of the chiral symmetry breaking in the QCD vacuum as the external magnetic field strengthens. The QCD vacuum can also acquire electromagnetic superconducting properties supported by condensation of electrically charged mesonic bound states with vector, $\rho$-meson quantum numbers~\cite{Chernodub:2010qx}. The transient magnetic fields of relevant scales appear in non-central and ultra-peripheral heavy-ion collisions at RHIC and LHC facilities~\cite{Skokov:2009qp, Deng:2012pc}. 

Electroweak fundamental interactions provide us with an additional source of vacuum instability at the critical magnetic field~\cite{Nielsen:1978rm, Skalozub1978, Ambjorn:1988fx}:
\beqn
B_{c1} \equiv B^{\mathrm{EW}}_{c1} = \frac{m_W^2}{e} \simeq 1.1 \times 10^{20}\,\mbox{T}\,,
\label{eq:eBc1}
\eeqn
determined by the mass $m_W \simeq 80.4\GeV$ of the $W$ boson. It was suggested that this instability marks the onset of the condensation of the $W$ bosons, which can be inferred from the classical equations of motion of the electroweak model~\cite{Skalozub1978, Skalozub:1986gw, Ambjorn:1988fx, Ambjorn:1988tm, Ambjorn:1988gb, Ambjorn:1989bd}. The condensate solution corresponds to a crystalline order of parallel vortex-like structures that share geometric similarity with the lattice of Abrikosov vortices of a conventional type-II superconductor: for realistically heavy Higgs masses, $m_H > m_Z$, the vortices in the $W$ condensate arrange themselves into a hexagonal lattice~\cite{Skalozub:1986gw, MacDowell:1991fw, Tornkvist:1992kh}. This exotic vacuum state should possess unusual anisotropic superconducting~\cite{Chernodub:2010qx} and superfluid~\cite{Chernodub:2012fi} properties~\footnote{The vacuum superconductivity at QCD~\cite{Chernodub:2010qx} and Electroweak~\cite{Chernodub:2010qx, Chernodub:2012fi} scales is similar to reentrant superconductivity which is suggested to occur in clean superconducting materials in very high magnetic fields~\cite{Rasolt1992}.}. The $W$ condensation may also develop in the cores of electroweak strings~\cite{Achucarro:1993bu, Perkins:1993qz, Olesen:1993ra, Garaud:2009uy}.

The electroweak vacuum is suggested to experience the second transition at an even higher magnetic field:
\beqn
B_{c2} \equiv B^{\mathrm{EW}}_{c2} = \frac{m_H^2}{e} \simeq 2.7 \times 10^{20}\,\mbox{T}\,,
\label{eq:eBc2}
\eeqn
determined by the Higgs mass $m_H = 125.1\,\mathrm{GeV}$. Above $B_{c2}$, the electroweak symmetry should be restored~\cite{Salam:1974xe,Linde:1975gx,Ambjorn:1989bd}. In this phase, the vortex lattice evaporates, leaving some traces in this new phase~\cite{Olesen:1991df, VanDoorsselaere:2012zb}. The magnetic fields of the relevant $10^{20}\,\mbox{T}$ scale might have been created at the cosmological electroweak phase transition in the first moments of the Early Universe~\cite{Vachaspati:1991nm, Grasso:2000wj}. Such enormous fields were suggested to exist even in the modern Universe in the vicinity of the magnetized black holes~\cite{Maldacena:2020skw, Ghosh:2020tdu}.

Our work aims to establish, using the first-principle lattice simulations, the phase structure of the vacuum subjected to magnetic fields of the electroweak strength. Despite the "weak" name, such fields are among the most powerful magnetic fields that were rarely discussed in the context of the Standard Model of particles.

The discussions of the effect of magnetic fields on the vacuum structure reveal certain controversies in the literature. The transition to the inhomogeneous superconducting phase of the electroweak (EW) vacuum proceeds via the instability of the vacuum at the first critical field~\eq{eq:eBc1} because at $B > B_{c1}$, the ground state $W$ mass becomes a purely imaginary quantity, $m_W^2(B) = m_W^2 - |e B|$. At the classical level, the formation of the periodic vortex lattice in the background magnetic field has been established in the EW model~\cite{Ho:2020ltr}. However, this classical-level scenario, together with the arguments based on loop computations~\cite{Nielsen:1978rm}, has been questioned in Ref.~\cite{Skalozub:2014epa} where it was shown that quantum corrections could add a radiative term to the classical $W$ mass in such a way that the mass does not vanish at the critical field $B = B_{c1}$. Consequently, it was concluded that no thermodynamic instability should occur in the Electroweak sector. Earlier numerical simulations of the electroweak model in the background magnetic field did not reveal the presence of the vortex-dominated phase around the finite-temperature electroweak crossover~\cite{Kajantie:1998rz}, which could be explained by a destructive role of thermal fluctuations.

A similar no-go theorem was suggested to forbid the superconducting transition in QCD vacuum~\cite{Hidaka:2012mz}. The instability in QCD should proceed via the spontaneous $\rho$-meson condensation similar to the magnetic-field induced condensation of the $W$ bosons in the EW model~\cite{Chernodub:2010qx}. The fact that the $\rho$-meson mass does not vanish at any magnetic field was later supported by the effective model calculations~\cite{Andreichikov:2013zba} as well as the first-principle numerical simulations~\cite{Bali:2017ian}. However, despite the absence of the thermodynamic singularity at finite magnetic field $B$, it was argued that the large-$B$ superconducting phase can still emerge via a smooth crossover transition implying that the transition to the new phase occurs at nonvanishing $\rho$-meson mass in the absence of a thermodynamic singularity~\cite{Chernodub:2013uja}. The latter scenario has a speculative nature that requires confirmation from a first-principle simulation. To this end, the electroweak model provides us with an exciting playground, given the similarity of the superconducting mechanisms in both systems.

\vskip 1mm 
\paragraph*{\bf Electroweak model.} 
We consider the bosonic sector of the Electroweak model with the Lagrangian
\beqn
\cL_\EW & = & -\frac{1}{2} \tr (W_{\mu \nu} W^{\mu \nu}) - \frac{1}{4} Y_{\mu \nu} Y^{\mu \nu} \nonumber \\
& & + (D_\mu \phi)^\dagger (D^\mu \phi) - V(\phi),
\label{eq:LEW}
\eeqn
where the field strengths of, respectively, the SU(2) gauge field $W_\mu^a$ and $U(1)_Y$ hypercharge gauge field $Y_\mu$ are
\beqn
W_{\mu \nu}^a & = & \partial_\mu W_\nu^a - \partial_\nu W_\mu^a + i g \varepsilon^{abc} W_\mu^b W_\nu^c\,, 
\label{eq:W:munu}\\
Y_{\mu \nu} & = & \partial_\mu Y_\nu - \partial_\nu Y_\mu\,,
\label{eq:Y:munu}
\eeqn
These vector fields interact with the complex scalar Higgs doublet $\phi \equiv (\phi_1,\phi_2)^T$ via the covariant derivative:
\beqn
    D_\mu = \partial_\mu + \frac{i}{2} g W_\mu^a \sigma^a + \frac{i}{2} g' Y_\mu,
\label{eq:D:mu}
\eeqn
where $\sigma^a$ ($a=1,2,3$) are the Pauli matrices. The ratio of the $U(1)$ and $SU(2)$ gauge couplings, $g'/g = \tan \theta_W$, defines the electroweak mixing (Weinberg) angle $\theta_W$ fixed in experiments~\cite{CODATA2018}:
$\sin^2 \theta_W \equiv 1 - m_W^2/m_Z^2 = 0.22290(30)$\,.

The last term in the Lagrangian~\eq{eq:LEW} is the potential $V(\phi) = \lambda \left(\phi^\dagger \phi - v^2/2 \right)^2$ of the Higgs field doublet $\phi$, where $\lambda$ is the dimensionless self-coupling of the Higgs field and the only dimensionful parameter $v$ sets the vacuum expectation value of the Higgs field. The only dimensionful parameter $v$ sets the overall mass scale in the system and gives, at low temperature, the vacuum expectation value to the Higgs field, $\avr{\phi} \neq 0$, which breaks the electroweak symmetry down to its electromagnetic subgroup, $SU(2)_W\times U(1)_Y \to U(1)_{\mathrm{e.m.}}$.

In the broken phase, the Higgs field acquires the mass $m_H = \sqrt{2 \lambda} v$. The theory possesses the massless photon,
\beqn
A_\mu = W_\mu^3 \sin \theta_W + Y_\mu \cos \theta_W\,,
\label{eq:A:field}
\eeqn
and three massive gauge bosons, which include the electrically (off-diagonal) charged $W$ bosons $W^\pm_\mu = W^1_\mu \pm i W^2_\mu$, and the neutral (diagonal) $Z$ boson:
\beqn
Z_\mu = W_\mu^3 \cos \theta_W - Y_\mu \sin \theta_W\,,
\label{eq:Z:field}
\eeqn
with the masses $m_W = g v/2$ and $m_Z = m_W/\cos \theta_W$.

\vskip 1mm 
\paragraph*{\bf Hypermagnetic field.} 
We consider the electroweak vacuum in the background of the hypermagnetic field ${\bs B}_Y = {\bs \nabla} \times {\bs Y}$ corresponding to the hypergauge field $Y^\mu = (Y^0, {\bs Y})$. In the broken phase, the two fields are related to each other:
\beqn
g' {\bs B}_Y = e {\bs B} \qquad {\mbox{[broken phase]}}\,,
\label{eq:BYB}
\eeqn
as it follows from the definition of the elementary electric charge, $e = g \sin\theta_W = g' \cos\theta_W = g g'/\sqrt{g^2 + g^{\prime 2}}$, Eqs.~\eq{eq:A:field} and \eq{eq:Z:field}, as well as from the fact that in the broken phase, the $Z$ boson is a massive particle which carries no global flux. In the symmetry-restored phase, where the magnetic field $\bs B$ cannot be defined, the hypermagnetic field ${\bs B}_Y$ plays the role of a genuine field.

\vskip 1mm 
\paragraph*{\bf Structure of electroweak vacuum in the magnetic field.} 
Using the first-principle Monte Carlo techniques, we simulate the lattice version of the EW model~\eq{eq:LEW}. The standard lattice discretization of the model, the known particularities of the lattice (hyper)magnetic field, the technicalities related to the choice of lattice parameters close to the continuum limit, and the lattice form of the physical observables discussed in the paper are described in the Supplemental Material, Sections A, B, C, and D respectively, which includes Refs.~\cite{Bunk:1992xt,Bunk:1992kf,Fodor:1994sj,Fodor:1994dm,Csikor:1996sp,Aoki:1999fi,Sakharov:1967dj,Rubakov:1996vz,Kajantie:1993ag,Ilgenfritz:1995sh,Gurtler:1996wx,Kajantie:1996qd,Kajantie:1996mn,Kajantie:1996qd,DOnofrio:2015gop,Langguth:1985dr,Bali:2011qj,Gattringer:2010zz,Bali:2011qj,
Frohlich:1980gj,Frohlich:1981yi,Woloshyn:2017rhe,Lewis:2018srt,Maas:2017wzi,Langguth:1985dr,Durr:2004xu,Langguth:1985dr,Bunk:1992xt,Fodor:1994dm,PDG,Csikor:1998eu,DOnofrio:2015gop,Csikor:1996sp,Osterwalder:1977pc,Fradkin:1978dv,Banks:1979fi,Seiler:2015rwa,Lee:1972fj,Caudy:2007sf}. 
Important subtleties of definitions of fields and particle contents in the standard Electroweak model and their lattice realizations are reviewed in Ref.~\cite{Maas:2017wzi}.

\begin{figure*}[!htb]
\begin{center}
\begin{tabular}{cc}
\includegraphics[width=0.47\textwidth,clip=true]{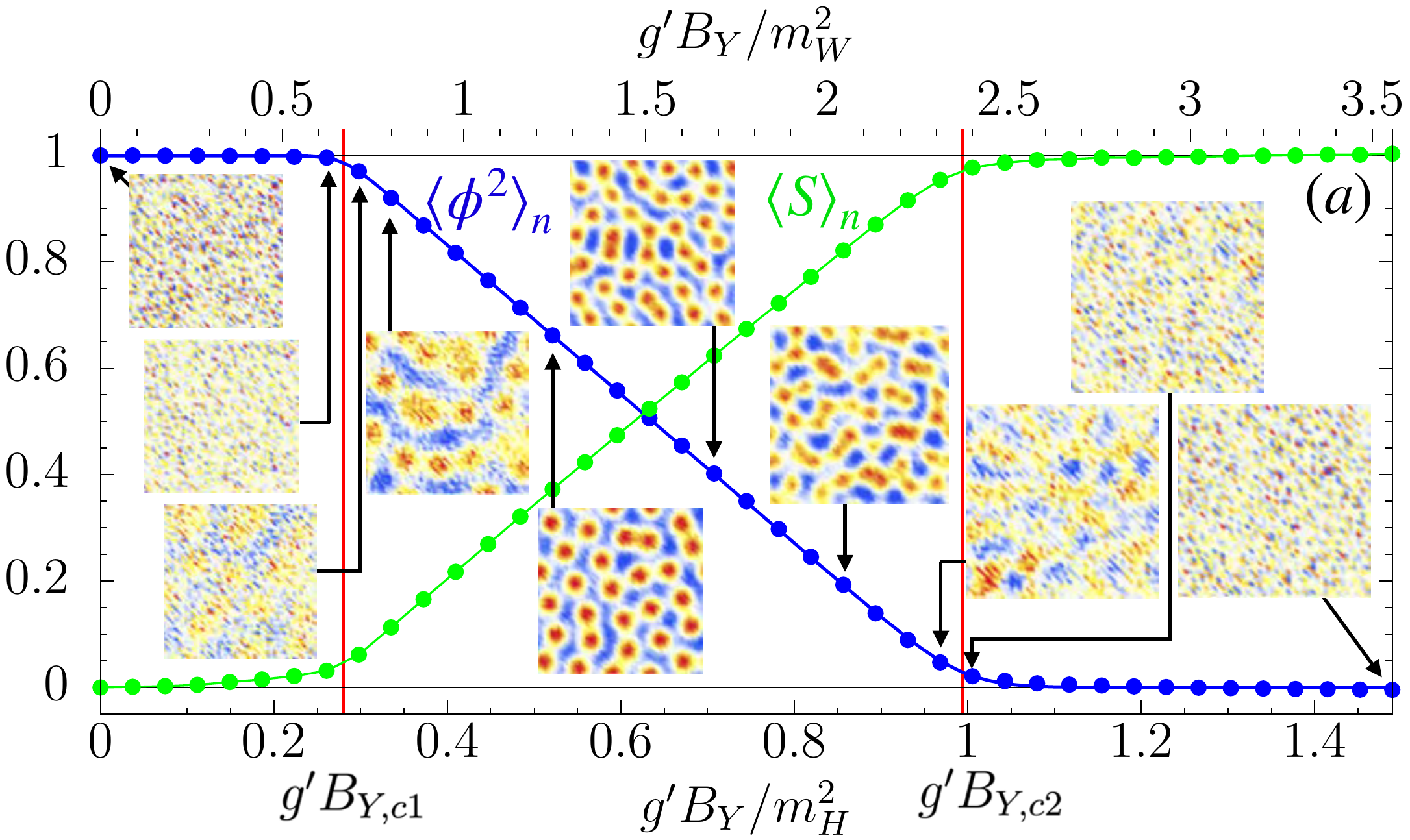} &
\includegraphics[width=0.49\textwidth,clip=true]{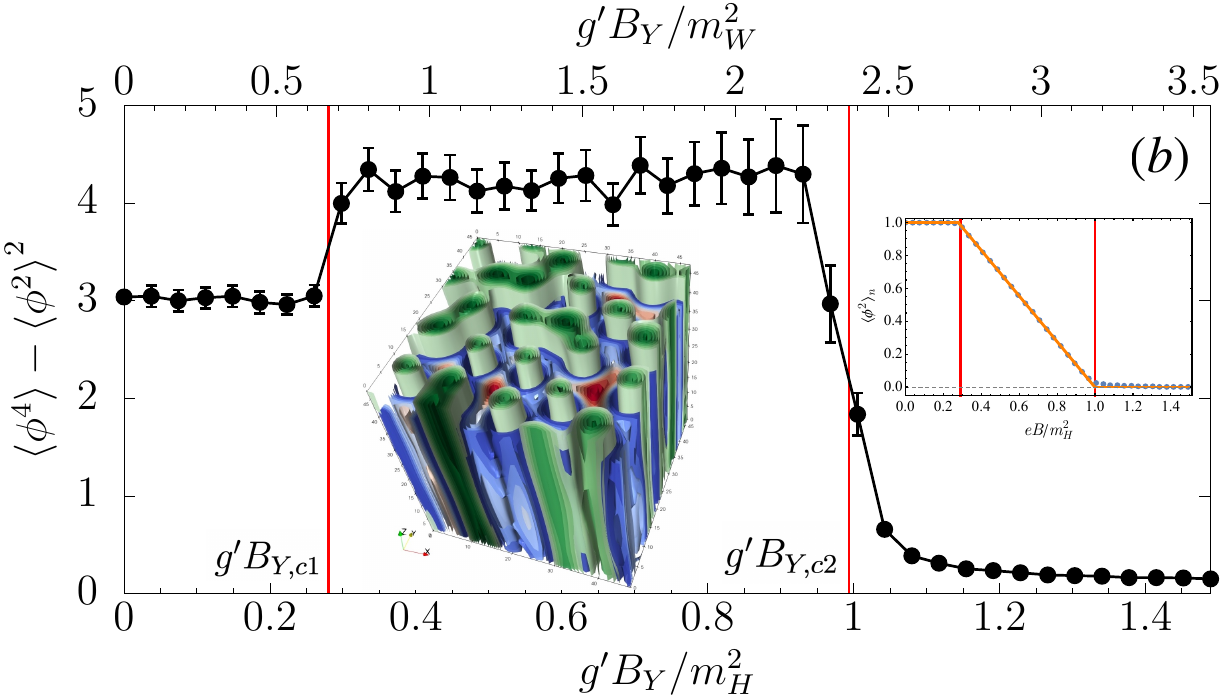}
\end{tabular}
\end{center}
\vskip -4mm 
\caption{(a) Normalized value $\langle\phi^2\rangle_n = \phi_r^2(B_Y)/\phi_r^2(0)$ of the additively renormalized, volume-averaged, $\phi^2 \equiv \frac{1}{V} \int \phi^\dagger \phi$, squared Higgs condensate $\phi_r^2(B_Y) = \vev{\phi^2(B_Y)} - \vev{\phi^2(\infty)}$ (blue) and the likewise normalized action $\langle S \rangle_n$ (green). The insets show the density plots of the $Z_{12}$ fluxes in the cross-sections normal to the magnetic field axis in typical configurations (more details are provided in the description of Fig.~\ref{fig:structure}). (b) The susceptibility of the Higgs field squared vs. the hypermagnetic field~$B_Y$. The right inset shows the fit of the Higgs condensate shown in (a) by the piecewise function~\eq{eq:phi2:piecewise}. The left inset illustrates a typical 3d configuration in the inhomogeneous phase in the (hyper)magnetic field background $g'  B_Y = e B = 1.1 m_W^2$ (the total number of vortices is 24). The equipotential surfaces of the $W$ condensate (the Higgs condensate) are shown in blue and red (green). These quantities, which take their maximal values at the centers of the corresponding structures, are shown in complementary regions. We used a cooling procedure to improve the visibility of this 3d picture. The vertical red lines denote the transitions.}
\label{fig:PhiSQ}
\end{figure*}

In Fig.~\ref{fig:PhiSQ}(a) we show the (normalized) vacuum expectation values of the Higgs field squared $\avr{|\phi|^2}$ and the action of the model -- the EW Lagrangian~\eq{eq:LEW} integrated over the whole spacetime -- as the functions of the background (hyper)magnetic field $g' B_Y$. These observables point out the existence of three phases separated by two pseudocritical magnetic fields:
\beqn
g' B_{Y,c1} \equiv e B_{c1} & = & 0.68(5) m_W^2\,, 
\label{eq:Bc1:lat} \\ 
g' B_{Y,c2} \equiv e B_{c2} & = & 0.99(2)m_H^2\,,
\label{eq:Bc2:lat}
\eeqn
identified as the (pairwise coinciding) inflection points of the Higgs condensate and the action, supporting the gauge-invariant nature of the transitions.

The first critical field~\eq{eq:Bc1:lat} turns out to be about 30\% weaker than value~\eq{eq:eBc1} predicted by the classical theoretical analysis that does not take into account quantum fluctuations. However, the second critical field~\eq{eq:Bc2:lat} agrees precisely with the theoretical value~\eq{eq:eBc2}. 

The classical picture predicts that the magnetic field affects the Higgs condensate as follows~\cite{Nielsen:1978rm,Skalozub1978,Skalozub:1986gw,Ambjorn:1988fx,Ambjorn:1988tm,Ambjorn:1988gb,Ambjorn:1989bd,MacDowell:1991fw,Salam:1974xe,Linde:1975gx}: 
\begin{itemize}
\item[(i)] In the broken phase ($B < B_{c1}$), the Higgs condensate does not depend on the magnetic field $B$.
\item[(ii)] When $B$ exceeds the first critical value $B = B_{c1}$, the vacuum develops a raising $W$ condensate which gradually inhibits the Higgs condensate.
\item[(iii)] Finally, as the field reaches the second critical value, $B = B_{c2}$, the Higgs condensate should vanish, and the electroweak theory should be restored.
\end{itemize}
All these properties are spectacularly confirmed 
in Fig.~\ref{fig:PhiSQ}(a), where the fluctuations of scalar excitations over the condensate are removed by the normalization~\footnote{Strictly speaking, the vanishing of the Higgs condensate at $B>B_{c2}$ is not evident from the normalized condensate squared of Fig.~\ref{fig:PhiSQ}(a) without additional analysis. Indeed, the quantity $\avr{\phi^2}$ is affected by significant ultraviolet contributions, which are always present in broken and unbroken phases. However, the homogenization of the $W$ condensate observed above $B_{c2}$ suggests that the Higgs condensate should be absent at $B > B_{c2}$.}.

The observed dependence of the Higgs expectation value on the magnetic field, shown in Fig.~\ref{fig:PhiSQ}(a), can be described by an impressively simple piecewise-linear formula predicted by the theory~\cite{Ambjorn:1989bd,Ambjorn:1989sz,Ambjorn:1992ca,Chernodub:2012fi}:
\beqn
\frac{\avr{\phi^2}_r(B)}{\avr{\phi^2}_r(0)} = 
\left\{
\begin{array}{lcl}
1, & \quad & B < B_{c1}, \\
\frac{B_{c2} - B}{B_{c2} - B_{c1}}, & \quad & B_{c1} < B < B_{c2}, \\
0, & \quad & B > B_{c2},
\end{array}
\right.
\label{eq:phi2:piecewise}
\eeqn
which fits our data everywhere except for small regions around the (pseudo)critical points $B = B_{c1}$ and $B = B_{c2}$. 

\begin{figure}[!htb]
\begin{center}
\includegraphics[width=0.485\textwidth,clip=true]{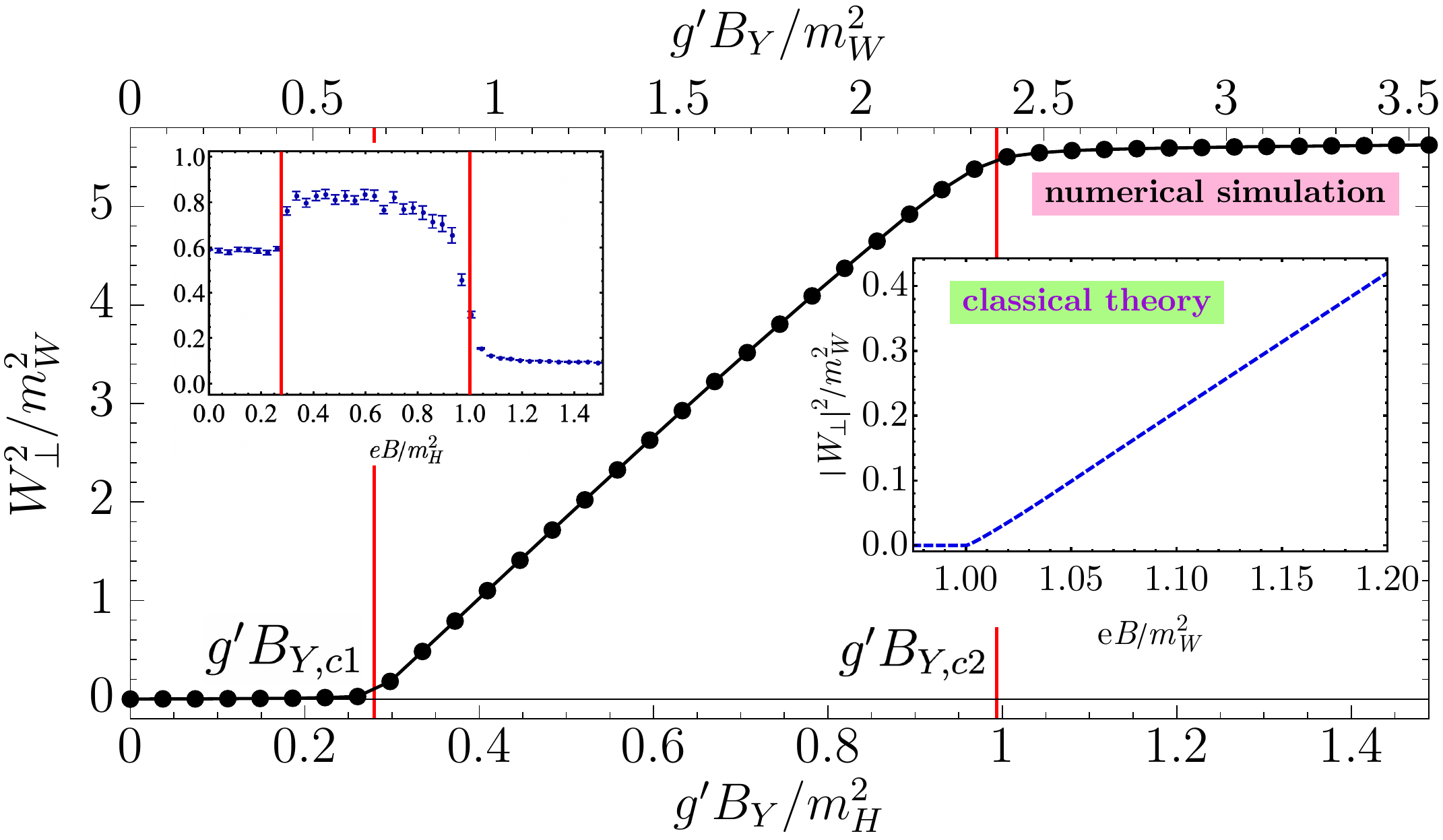}
\end{center}
\vskip -4mm 
\caption{Expectation value of the $W$ condensate vs. the (hyper)magnetic field $g' B_Y = e B$. The right inset is the theoretical result based on the classical solution for the transverse $W$ condensate squared~\cite{Chernodub:2012fi} around the first transition. The left inset shows the susceptibility of the transverse $W$ condensate.}
\label{fig:W2}
\end{figure}

The structure of the classical solution around $B_{c1}$ implies that the first phase transition should be of the second order~\cite{Ambjorn:1988gb,Chernodub:2012fi}. In this case, the susceptibility of the Higgs field should possess a local maximum at the (pseudo)critical point. We do not see any peaks in the susceptibility across either $B_{c1}$ or $B_{c2}$, Fig.~\ref{fig:PhiSQ}(b). Thus, these transitions are smooth crossovers. 

\begin{figure*}[!thb]
\includegraphics[width=1\linewidth]{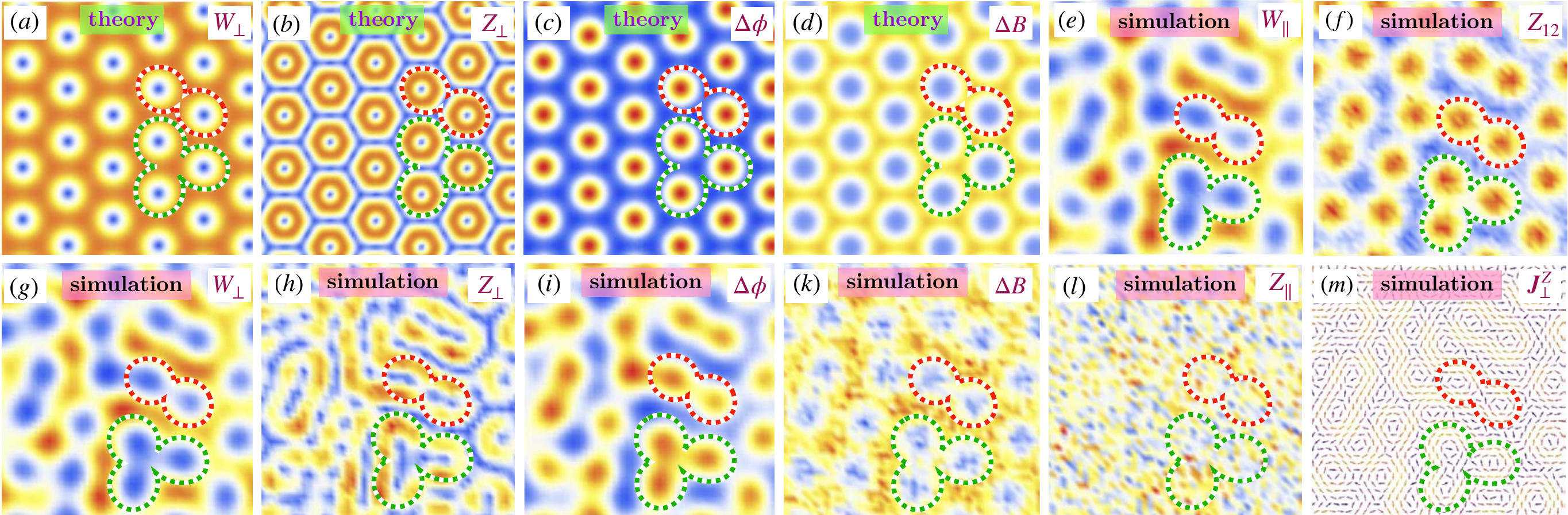}
\caption{Density plots of various quantities in the cross-sections normal to the axis of the (hyper)magnetic field. 
Theoretical results, (a)-(d), are given for the classical solution of Ref.~\cite{Chernodub:2012fi} at $B = 1.01 B_c$. The numerical results of the first-principle simulations, (e)-(m), are given for a typical lattice configuration in the background of the (hyper)magnetic field $g' B_Y \equiv e B \simeq 1.1 m_W^2 \simeq 1.6 e B_{c1}$ (the magnetic number $k=12$ for our lattices). 
(a) and (g): transverse $W$ condensate $W_\perp = \sqrt{|W_x|^2 + |W_y|^2}$; 
(b) and (h):  transverse $Z$ condensate $Z_\perp = \sqrt{|Z_x|^2 + |Z_y|^2}$; 
(c) and (i) local excess of the Higgs expectation value over the condensate, $\Delta \phi({\bs x}) = \phi({\bs x}) - \avr{\phi}$;
(d) and (k): local excess of the magnetic field value over the background,  $\Delta B({\bs x})  = B({\bs x})  - B_{\mathrm{ext}}$;
(e) and (l): longitudinal $W$ and $Z$ condensates, $W_\| = |W_z|$ and $Z_\| = |Z_z|$, respectively (theoretically, $W_\| = Z_\| = 0$ at the classical level);
(f): the $Z$-flux; 
(m) the neutral Higgs currents ${\bs J}_\perp^{Z}$.
The red (blue) colors correspond to maxima (minima); absolute values are given for complex quantities. 
The same regions are circumvented by the red and green dashed lines (separately for the analytical solution and simulated configuration) to guide the eye.
Each simulated picture is obtained by averaging data over s part of the Markov chain (last $5000$ configurations). Averaging over the entire ensemble leads to the blurring of the vortex structure.}
\label{fig:structure}
\end{figure*}

The $W$ condensates are shown in Fig.~\ref{fig:W2}. In an excellent qualitative agreement with the theory (the right inset), the squared $W_\perp^2$ condensate raises linearly in the intermediate phase. The observed slope of the linear part, $\partial |W_\perp|^2/\partial (eB) \simeq 2.9$, is about 30\% larger than the slope predicted by the classical solution, $\partial |W_\perp|^2/\partial (eB) \simeq 2.1$ ~\cite{Chernodub:2012fi}. This deviation indicates the important role of the quantum fluctuations responsible also for the 30\% shift of the first critical field $B_{c1}$. In the restored phase at $B > B_{c2}$, the $W$ condensate flattens, possibly indicating the presence of the condensate of ``zero-field twists'' which are suggested to be the remnants of the vortex lattice~\cite{Olesen:1991df} visible close to $B = B_{c2}$~\cite{VanDoorsselaere:2012zb}. The $W$ susceptibility (the left inset) exhibits close similarity with the susceptibility of the Higgs field shown in Fig.~\ref{fig:PhiSQ}(b).

\vskip 1mm 
\paragraph*{\bf Nature of the intermediate inhomogeneous phase.}  To confront our theoretical expectation with the first principle simulations, we visualize in Fig.~\ref{fig:structure} the structure of the electroweak fields in the cross-section perpendicular to the magnetic field axis (we take $\bs B$ in the $z$ direction). We show analytical results in the classical theory in Figs.~\ref{fig:structure}(a)-(d) and visualize the numerical data obtained in lattice simulations in Figs.~\ref{fig:structure}(e)-(m). The numerical results were obtained by taking an average over a few dozen successive field configurations generated in the background of the (hyper)magnetic field $g' B_Y \equiv e B \simeq 1.1 m_W^2$ which corresponds to the intermediate phase in between two critical fields $B_{c1} < B < B_{c2}$. 

According to the theoretical expectations~\cite{Skalozub:1986gw,Ambjorn:1989bd,MacDowell:1991fw}, the ground state of the intermediate phase corresponds to a spatially inhomogeneous structure made of the $W$ condensate with nonvanishing transverse components $W_x$ and $W_y$. The inhomogeneities are produced by vortices that are embedded in the condensate. The vortices should arrange themselves into a hexagonal pattern in the plane perpendicular to the magnetic field. The transverse component of the $W$ condensate, $W_\perp = \sqrt{|W_x|^2 + |W_y|^2}$, should vanish in the core of each vortex, Fig.~\ref{fig:structure}(a).

Instead of the hexagonal pattern of the classical solution, Fig.~\ref{fig:structure}(a), our lattice simulations reveal a less regular structure, Fig.~\ref{fig:structure}(g). However, similarly to the classical solution, the lattice field $W_\perp$ exhibits a semi-classical behavior characterized by large, both in magnitude and in size structures~\footnote{We do not perform any smoothening or cooling procedures that could drive our Monte Carlo configurations of Figs.~\ref{fig:structure}(e)-(m) closer to the classical regime.}. We associate these structures with the inhomogeneous $W$ condensate. The condensate exhibits a set of separate deep minima that point to the presence of the vortex cores in agreement with the theoretical classical picture, Fig.~\ref{fig:structure}(a). Modulo occasional overlaps, the total number of vortices at the chosen magnetic number $k=12$ appears to be equal to 24, which corresponds to the number $2k$ of the elementary fluxes of the hypermagnetic field, as expected.

Contrary to the expectations based on the classical theory, Fig.~\ref{fig:structure}(a), the vortices do not form the crystalline phase in the vacuum, Fig.~\ref{fig:structure}(g). While some traces of the crystalline vortex order are seen, the quantum fluctuations disorder the classical hexagonal structure so that the vortices form a disordered solid or, possibly, a liquid. The formation of the vortex liquid phase is not unexpected, though, as it has been proposed in a similar non-Abelian context in Ref.~\cite{Ambjorn:1979xi}.

According to the classical picture, the transverse $Z$ condensate $Z_\perp = \sqrt{|Z_x|^2 + |Z_y|^2}$ forms a regular honeycomb structure, Fig.~\ref{fig:structure}(b). This neutral condensate vanishes in the center of each vortex, and a honeycomb-like manifold is in between the vortices. The regions with nonzero $Z$ condensate are thin pipe-like shells surrounding the vortex cores. Strikingly, these classical structures, disordered by quantum fluctuations, are also seen in our lattice configurations, Fig.~\ref{fig:structure}(h): the thin shells of the $Z$ condensate surround the cores of vortices.

The presence of classically large $W$ and $Z$ condensates implies that the vacuum of the electroweak model in the intermediate phase becomes a highly anisotropic electromagnetic superconductor that exhibits dissipationless transport electric charge accompanied by frictionless flow of a neutral superfluid component~\cite{Chernodub:2010qx,Chernodub:2012fi}. The formation of the vacuum superconductivity and the crystalline vortex phase is substantially different from the vortex lattices observed, for example, in the mixed Abrikosov phase of a type~II superconductor.

The classical EW theory predicts that in the cores of vortices, the Higgs condensate should get enhanced~\cite{Chernodub:2012fi}, Fig.~\ref{fig:structure}(c), while the magnetic field should be locally suppressed due to the anti-screening effect~\cite{Ambjorn:1988fx}, Fig.~\ref{fig:structure}(d). These properties, which defy our intuition based on the Abrikosov picture of type-II superconductors, are confirmed by the results of our numerical simulations shown in Figs.~\ref{fig:structure}(i) and (k), respectively. A 3d picture of the Higgs and $W$ condensates and the magnetic field lines of a typical configuration is shown in the inset of Fig.~\ref{fig:PhiSQ}(b).

We also noticed that $|W_x| \simeq |W_y|$ holds high precision in numerical simulations. The unexpected outcome of our simulations is the presence of the large longitudinal condensate $W_\| \simeq W_\perp$, Fig.~\ref{fig:structure}(e) which closely mimics the transverse condensate, Fig.~\ref{fig:structure}(g). This observation disagrees with the classical theory that predicts $W_\| = 0$ in the ground state. On the contrary, the condensate of the $z$ component of the neutral $Z$ boson is vanishing in agreement with the classical picture: we observe only small quantum fluctuations in this quantity, Fig.~\ref{fig:structure}(l).

Numerically, the $Z_{12}$ flux provides us with the most transparent view of the vortex content of field configurations, Fig.~\ref{fig:structure}(f). The peaks in the $Z$ flux point out to the positions of the vortex cores also seen as the deeps in the $W$ condensate, Fig.~\ref{fig:structure}(e), the spikes in the Higgs condensate, Fig.~\ref{fig:structure}(g), and, much less clear, as the minima in the magnetic field, Fig.~\ref{fig:structure}(h). The associated neutral currents ${\bs J}^Z$ of the Higgs field, Fig.~\ref{fig:structure}(m), defined as a variation of the matter part of the action with respect to the $Z$ field, circumvents the vortices.

In the inset of Fig.~\ref{fig:PhiSQ}(a), we show the evolution of the $Z$ flux density in the transverse plane of the gradually increasing (hyper)magnetic field. The vortices start to form as soon as the magnetic field crosses the first pseudocritical value, $B = B_{c1}$. The vortex structures are barely seen. The fuzziness of vortex positions appears due to the weakness of the condensates right above the critical point. This property makes the weak classical structure vulnerable to the disorder caused by ultraviolet fluctuations and phonons in the vortex lattice that lead to the drifting of the vortex cores. The vortices may form a liquid close to the first critical field $B_{c1}$.

In the middle of the superconducting phase, the vortex liquid partially solidifies into a disordered solid. The physical motion of the vortices leads to enhanced local fluctuations of all physical quantities that experience extrema at or around vortex cores. In particular, the vortex motion enhances fluctuations of the Higgs condensate, thus leading to the elevated values of the Higgs and $W$ susceptibilities in the inhomogeneous phase that we already observed in Fig.~\ref{fig:PhiSQ}(b) and Fig.~\ref{fig:W2}, respectively. 

Close to the second critical field, $B = B_{c2}$, the vortex solid starts to melt. Finally, the vortices disappear entirely as the vacuum crosses into the third phase at $B > B_{c2}$, where the electroweak symmetry gets restored.

\vskip 1mm
\paragraph{\bf Conclusions.}  Using first-principle numerical simulations, we establish the three-phase structure of the electroweak sector of the vacuum in the background of the strong magnetic field at zero temperature. In agreement with the theoretical analysis~\cite{Nielsen:1978rm,Skalozub1978,Skalozub:1986gw,Ambjorn:1988fx,Ambjorn:1988tm,Ambjorn:1988gb,Ambjorn:1989bd,MacDowell:1991fw,Salam:1974xe,Linde:1975gx}, we find that the vacuum experiences two successive transitions. Contrary to expectations, these transitions are smooth crossovers.

As the magnetic field reaches the first pseudocritical value~\eq{eq:Bc1:lat}, the vacuum turns into the intermediate inhomogeneous phase, characterized by the presence of classically large $W$ and $Z$ condensates. Then, at the second pseudocritical field \eq{eq:Bc2:lat}, the inhomogeneities disappear, and the electroweak symmetry gets restored. The inhomogeneous phase is populated by vortices that locally have an almost classical field structure, mostly in agreement with Refs.~\cite{Ambjorn:1988gb,Chernodub:2012fi}. The classical hexagonal order of the vortex phase is, however, not realized due to strong quantum fluctuations: we find the evidence that in the middle of the inhomogeneous phase, the vortices appear to form a disordered vortex solid which melts, closer to both pseudo-critical magnetic fields, into a vortex liquid.

Summarising, we have shown in first-principle calculations that the electroweak sector of the vacuum experiences two consecutive crossover transitions in strengthening the magnetic field background. At intermediate fields in between these crossovers, we observe the appearance of large, inhomogeneous structures consistent with theoretically expected~\cite{Skalozub1978,Skalozub:1986gw,Ambjorn:1988fx,Ambjorn:1988tm,Ambjorn:1988gb,Ambjorn:1989bd} a classical picture of the formation of $W$ and $Z$ condensates pierced by vortices that form a disordered solid or a liquid rather than an expected solid. The presence of classically significant $W$ and $Z$ condensates points to the fascinating possibility that in the strong magnetic field, the vacuum becomes an electromagnetic superconductor enriched by a neutral superfluid component that support dissipationless transport along magnetic field lines~\cite{Chernodub:2010qx,Chernodub:2012fi}. In the present time, such conditions can be realized in the vicinity of the magnetized black holes~\cite{Maldacena:2020skw,Ghosh:2020tdu}.

\begin{acknowledgments}
\paragraph*{\bf Acknowledgments.}  
M.C. is grateful to M.~Shaposhnikov for illuminating discussions. The numerical simulations were performed at the Supercomputer SQUID (Osaka University, Japan) and the computing cluster Vostok-1 of Far Eastern Federal University. V.G. has been supported by RSF (Project No. 21-72-00121, \href{https://rscf.ru/project/21-72-00121/}{https://rscf.ru/project/21-72-00121/} devoted to a code development, simulation, and data analysis). A.M. has been partially supported by the state assignment of the Ministry of Science and Higher Education of Russia (Project No. 0657-2020-0015). M.C. has been partially supported by the project IEA-International Emerging Actions No. 00677.
\end{acknowledgments}


\begin{thebibliography}{79}%
\makeatletter
\providecommand \@ifxundefined [1]{%
 \@ifx{#1\undefined}
}%
\providecommand \@ifnum [1]{%
 \ifnum #1\expandafter \@firstoftwo
 \else \expandafter \@secondoftwo
 \fi
}%
\providecommand \@ifx [1]{%
 \ifx #1\expandafter \@firstoftwo
 \else \expandafter \@secondoftwo
 \fi
}%
\providecommand \natexlab [1]{#1}%
\providecommand \enquote  [1]{``#1''}%
\providecommand \bibnamefont  [1]{#1}%
\providecommand \bibfnamefont [1]{#1}%
\providecommand \citenamefont [1]{#1}%
\providecommand \href@noop [0]{\@secondoftwo}%
\providecommand \href [0]{\begingroup \@sanitize@url \@href}%
\providecommand \@href[1]{\@@startlink{#1}\@@href}%
\providecommand \@@href[1]{\endgroup#1\@@endlink}%
\providecommand \@sanitize@url [0]{\catcode `\\12\catcode `\$12\catcode
  `\&12\catcode `\#12\catcode `\^12\catcode `\_12\catcode `\%12\relax}%
\providecommand \@@startlink[1]{}%
\providecommand \@@endlink[0]{}%
\providecommand \url  [0]{\begingroup\@sanitize@url \@url }%
\providecommand \@url [1]{\endgroup\@href {#1}{\urlprefix }}%
\providecommand \urlprefix  [0]{URL }%
\providecommand \Eprint [0]{\href }%
\providecommand \doibase [0]{http://dx.doi.org/}%
\providecommand \selectlanguage [0]{\@gobble}%
\providecommand \bibinfo  [0]{\@secondoftwo}%
\providecommand \bibfield  [0]{\@secondoftwo}%
\providecommand \translation [1]{[#1]}%
\providecommand \BibitemOpen [0]{}%
\providecommand \bibitemStop [0]{}%
\providecommand \bibitemNoStop [0]{.\EOS\space}%
\providecommand \EOS [0]{\spacefactor3000\relax}%
\providecommand \BibitemShut  [1]{\csname bibitem#1\endcsname}%
\let\auto@bib@innerbib\@empty
\bibitem [{Note1()}]{Note1}%
  \BibitemOpen
  \bibinfo {note} {We use the units $\hbar = c = 1$.}\BibitemShut {Stop}%
\bibitem [{\citenamefont {Schwinger}(1951)}]{Schwinger:1951nm}%
  \BibitemOpen
  \bibfield  {author} {\bibinfo {author} {\bibfnamefont {J.~S.}\ \bibnamefont
  {Schwinger}},\ }\href {\doibase 10.1103/PhysRev.82.664} {\bibfield  {journal}
  {\bibinfo  {journal} {Phys. Rev.}\ }\textbf {\bibinfo {volume} {82}},\
  \bibinfo {pages} {664} (\bibinfo {year} {1951})}\BibitemShut {NoStop}%
\bibitem [{\citenamefont {Olausen}\ and\ \citenamefont
  {Kaspi}(2014)}]{Olausen2014mcgill}%
  \BibitemOpen
  \bibfield  {author} {\bibinfo {author} {\bibfnamefont {S.}~\bibnamefont
  {Olausen}}\ and\ \bibinfo {author} {\bibfnamefont {V.}~\bibnamefont
  {Kaspi}},\ }\href@noop {} {\bibfield  {journal} {\bibinfo  {journal} {The
  Astrophysical Journal Supplement Series}\ }\textbf {\bibinfo {volume}
  {212}},\ \bibinfo {pages} {6} (\bibinfo {year} {2014})}\BibitemShut {NoStop}%
\bibitem [{\citenamefont {Adler}(1971)}]{Adler:1971wn}%
  \BibitemOpen
  \bibfield  {author} {\bibinfo {author} {\bibfnamefont {S.~L.}\ \bibnamefont
  {Adler}},\ }\href {\doibase 10.1016/0003-4916(71)90154-0} {\bibfield
  {journal} {\bibinfo  {journal} {Annals Phys.}\ }\textbf {\bibinfo {volume}
  {67}},\ \bibinfo {pages} {599} (\bibinfo {year} {1971})}\BibitemShut
  {NoStop}%
\bibitem [{\citenamefont {Shaviv}\ \emph {et~al.}(1999)\citenamefont {Shaviv},
  \citenamefont {Heyl},\ and\ \citenamefont {Lithwick}}]{Shaviv1999magnetic}%
  \BibitemOpen
  \bibfield  {author} {\bibinfo {author} {\bibfnamefont {N.~J.}\ \bibnamefont
  {Shaviv}}, \bibinfo {author} {\bibfnamefont {J.~S.}\ \bibnamefont {Heyl}}, \
  and\ \bibinfo {author} {\bibfnamefont {Y.}~\bibnamefont {Lithwick}},\
  }\href@noop {} {\bibfield  {journal} {\bibinfo  {journal} {Monthly Notices of
  the Royal Astronomical Society}\ }\textbf {\bibinfo {volume} {306}},\
  \bibinfo {pages} {333} (\bibinfo {year} {1999})}\BibitemShut {NoStop}%
\bibitem [{\citenamefont {Klevansky}\ and\ \citenamefont
  {Lemmer}(1989)}]{Klevansky:1989vi}%
  \BibitemOpen
  \bibfield  {author} {\bibinfo {author} {\bibfnamefont {S.~P.}\ \bibnamefont
  {Klevansky}}\ and\ \bibinfo {author} {\bibfnamefont {R.~H.}\ \bibnamefont
  {Lemmer}},\ }\href {\doibase 10.1103/PhysRevD.39.3478} {\bibfield  {journal}
  {\bibinfo  {journal} {Phys. Rev. D}\ }\textbf {\bibinfo {volume} {39}},\
  \bibinfo {pages} {3478} (\bibinfo {year} {1989})}\BibitemShut {NoStop}%
\bibitem [{\citenamefont {Klimenko}(1992)}]{Klimenko:1991he}%
  \BibitemOpen
  \bibfield  {author} {\bibinfo {author} {\bibfnamefont {K.~G.}\ \bibnamefont
  {Klimenko}},\ }\href {\doibase 10.1007/BF01566663} {\bibfield  {journal}
  {\bibinfo  {journal} {Z. Phys. C}\ }\textbf {\bibinfo {volume} {54}},\
  \bibinfo {pages} {323} (\bibinfo {year} {1992})}\BibitemShut {NoStop}%
\bibitem [{\citenamefont {Shovkovy}(2013)}]{Shovkovy:2012zn}%
  \BibitemOpen
  \bibfield  {author} {\bibinfo {author} {\bibfnamefont {I.~A.}\ \bibnamefont
  {Shovkovy}},\ }\href {\doibase 10.1007/978-3-642-37305-3_2} {\bibfield
  {journal} {\bibinfo  {journal} {Lect. Notes Phys.}\ }\textbf {\bibinfo
  {volume} {871}},\ \bibinfo {pages} {13} (\bibinfo {year} {2013})},\ \Eprint
  {http://arxiv.org/abs/1207.5081} {arXiv:1207.5081 [hep-ph]} \BibitemShut
  {NoStop}%
\bibitem [{\citenamefont {Chernodub}(2010)}]{Chernodub:2010qx}%
  \BibitemOpen
  \bibfield  {author} {\bibinfo {author} {\bibfnamefont {M.~N.}\ \bibnamefont
  {Chernodub}},\ }\href {\doibase 10.1103/PhysRevD.82.085011} {\bibfield
  {journal} {\bibinfo  {journal} {Phys. Rev. D}\ }\textbf {\bibinfo {volume}
  {82}},\ \bibinfo {pages} {085011} (\bibinfo {year} {2010})},\ \Eprint
  {http://arxiv.org/abs/1008.1055} {arXiv:1008.1055 [hep-ph]} \BibitemShut
  {NoStop}%
\bibitem [{\citenamefont {Skokov}\ \emph {et~al.}(2009)\citenamefont {Skokov},
  \citenamefont {Illarionov},\ and\ \citenamefont {Toneev}}]{Skokov:2009qp}%
  \BibitemOpen
  \bibfield  {author} {\bibinfo {author} {\bibfnamefont {V.}~\bibnamefont
  {Skokov}}, \bibinfo {author} {\bibfnamefont {A.~Y.}\ \bibnamefont
  {Illarionov}}, \ and\ \bibinfo {author} {\bibfnamefont {V.}~\bibnamefont
  {Toneev}},\ }\href {\doibase 10.1142/S0217751X09047570} {\bibfield  {journal}
  {\bibinfo  {journal} {Int. J. Mod. Phys. A}\ }\textbf {\bibinfo {volume}
  {24}},\ \bibinfo {pages} {5925} (\bibinfo {year} {2009})},\ \Eprint
  {http://arxiv.org/abs/0907.1396} {arXiv:0907.1396 [nucl-th]} \BibitemShut
  {NoStop}%
\bibitem [{\citenamefont {Deng}\ and\ \citenamefont
  {Huang}(2012)}]{Deng:2012pc}%
  \BibitemOpen
  \bibfield  {author} {\bibinfo {author} {\bibfnamefont {W.-T.}\ \bibnamefont
  {Deng}}\ and\ \bibinfo {author} {\bibfnamefont {X.-G.}\ \bibnamefont
  {Huang}},\ }\href {\doibase 10.1103/PhysRevC.85.044907} {\bibfield  {journal}
  {\bibinfo  {journal} {Phys. Rev. C}\ }\textbf {\bibinfo {volume} {85}},\
  \bibinfo {pages} {044907} (\bibinfo {year} {2012})},\ \Eprint
  {http://arxiv.org/abs/1201.5108} {arXiv:1201.5108 [nucl-th]} \BibitemShut
  {NoStop}%
\bibitem [{\citenamefont {Nielsen}\ and\ \citenamefont
  {Olesen}(1978)}]{Nielsen:1978rm}%
  \BibitemOpen
  \bibfield  {author} {\bibinfo {author} {\bibfnamefont {N.~K.}\ \bibnamefont
  {Nielsen}}\ and\ \bibinfo {author} {\bibfnamefont {P.}~\bibnamefont
  {Olesen}},\ }\href {\doibase 10.1016/0550-3213(78)90377-2} {\bibfield
  {journal} {\bibinfo  {journal} {Nucl. Phys. B}\ }\textbf {\bibinfo {volume}
  {144}},\ \bibinfo {pages} {376} (\bibinfo {year} {1978})}\BibitemShut
  {NoStop}%
\bibitem [{\citenamefont {Skalozub}(1978)}]{Skalozub1978}%
  \BibitemOpen
  \bibfield  {author} {\bibinfo {author} {\bibfnamefont {V.}~\bibnamefont
  {Skalozub}},\ }\href@noop {} {\bibfield  {journal} {\bibinfo  {journal} {Sov.
  J. Nucl. Phys.(Engl. Transl.);(United States)}\ }\textbf {\bibinfo {volume}
  {28}} (\bibinfo {year} {1978})}\BibitemShut {NoStop}%
\bibitem [{\citenamefont {Ambjorn}\ and\ \citenamefont
  {Olesen}(1988)}]{Ambjorn:1988fx}%
  \BibitemOpen
  \bibfield  {author} {\bibinfo {author} {\bibfnamefont {J.}~\bibnamefont
  {Ambjorn}}\ and\ \bibinfo {author} {\bibfnamefont {P.}~\bibnamefont
  {Olesen}},\ }\href {\doibase 10.1016/0370-2693(88)90120-7} {\bibfield
  {journal} {\bibinfo  {journal} {Phys. Lett. B}\ }\textbf {\bibinfo {volume}
  {214}},\ \bibinfo {pages} {565} (\bibinfo {year} {1988})}\BibitemShut
  {NoStop}%
\bibitem [{\citenamefont {Skalozub}(1987)}]{Skalozub:1986gw}%
  \BibitemOpen
  \bibfield  {author} {\bibinfo {author} {\bibfnamefont {V.~V.}\ \bibnamefont
  {Skalozub}},\ }\href@noop {} {\bibfield  {journal} {\bibinfo  {journal} {Sov.
  J. Nucl. Phys.}\ }\textbf {\bibinfo {volume} {45}},\ \bibinfo {pages} {1058}
  (\bibinfo {year} {1987})}\BibitemShut {NoStop}%
\bibitem [{\citenamefont {Ambjorn}\ and\ \citenamefont
  {Olesen}(1989{\natexlab{a}})}]{Ambjorn:1988tm}%
  \BibitemOpen
  \bibfield  {author} {\bibinfo {author} {\bibfnamefont {J.}~\bibnamefont
  {Ambjorn}}\ and\ \bibinfo {author} {\bibfnamefont {P.}~\bibnamefont
  {Olesen}},\ }\href {\doibase 10.1016/0550-3213(89)90004-7} {\bibfield
  {journal} {\bibinfo  {journal} {Nucl. Phys. B}\ }\textbf {\bibinfo {volume}
  {315}},\ \bibinfo {pages} {606} (\bibinfo {year}
  {1989}{\natexlab{a}})}\BibitemShut {NoStop}%
\bibitem [{\citenamefont {Ambjorn}\ and\ \citenamefont
  {Olesen}(1989{\natexlab{b}})}]{Ambjorn:1988gb}%
  \BibitemOpen
  \bibfield  {author} {\bibinfo {author} {\bibfnamefont {J.}~\bibnamefont
  {Ambjorn}}\ and\ \bibinfo {author} {\bibfnamefont {P.}~\bibnamefont
  {Olesen}},\ }\href {\doibase 10.1016/0370-2693(89)90476-0} {\bibfield
  {journal} {\bibinfo  {journal} {Phys. Lett. B}\ }\textbf {\bibinfo {volume}
  {218}},\ \bibinfo {pages} {67} (\bibinfo {year} {1989}{\natexlab{b}})},\
  \bibinfo {note} {[Erratum: Phys.Lett.B 220, 659 (1989)]}\BibitemShut
  {NoStop}%
\bibitem [{\citenamefont {Ambjorn}\ and\ \citenamefont
  {Olesen}(1990{\natexlab{a}})}]{Ambjorn:1989bd}%
  \BibitemOpen
  \bibfield  {author} {\bibinfo {author} {\bibfnamefont {J.}~\bibnamefont
  {Ambjorn}}\ and\ \bibinfo {author} {\bibfnamefont {P.}~\bibnamefont
  {Olesen}},\ }\href {\doibase 10.1016/0550-3213(90)90307-Y} {\bibfield
  {journal} {\bibinfo  {journal} {Nucl. Phys. B}\ }\textbf {\bibinfo {volume}
  {330}},\ \bibinfo {pages} {193} (\bibinfo {year}
  {1990}{\natexlab{a}})}\BibitemShut {NoStop}%
\bibitem [{\citenamefont {MacDowell}\ and\ \citenamefont
  {Tornkvist}(1992)}]{MacDowell:1991fw}%
  \BibitemOpen
  \bibfield  {author} {\bibinfo {author} {\bibfnamefont {S.~W.}\ \bibnamefont
  {MacDowell}}\ and\ \bibinfo {author} {\bibfnamefont {O.}~\bibnamefont
  {Tornkvist}},\ }\href {\doibase 10.1103/PhysRevD.45.3833} {\bibfield
  {journal} {\bibinfo  {journal} {Phys. Rev. D}\ }\textbf {\bibinfo {volume}
  {45}},\ \bibinfo {pages} {3833} (\bibinfo {year} {1992})}\BibitemShut
  {NoStop}%
\bibitem [{\citenamefont {Tornkvist}(1992)}]{Tornkvist:1992kh}%
  \BibitemOpen
  \bibfield  {author} {\bibinfo {author} {\bibfnamefont {O.}~\bibnamefont
  {Tornkvist}},\ }\href@noop {} {\  (\bibinfo {year} {1992})},\ \Eprint
  {http://arxiv.org/abs/hep-ph/9204235} {arXiv:hep-ph/9204235} \BibitemShut
  {NoStop}%
\bibitem [{\citenamefont {Chernodub}\ \emph {et~al.}(2013)\citenamefont
  {Chernodub}, \citenamefont {Van~Doorsselaere},\ and\ \citenamefont
  {Verschelde}}]{Chernodub:2012fi}%
  \BibitemOpen
  \bibfield  {author} {\bibinfo {author} {\bibfnamefont {M.~N.}\ \bibnamefont
  {Chernodub}}, \bibinfo {author} {\bibfnamefont {J.}~\bibnamefont
  {Van~Doorsselaere}}, \ and\ \bibinfo {author} {\bibfnamefont
  {H.}~\bibnamefont {Verschelde}},\ }\href {\doibase
  10.1103/PhysRevD.88.065006} {\bibfield  {journal} {\bibinfo  {journal} {Phys.
  Rev. D}\ }\textbf {\bibinfo {volume} {88}},\ \bibinfo {pages} {065006}
  (\bibinfo {year} {2013})},\ \Eprint {http://arxiv.org/abs/1203.5963}
  {arXiv:1203.5963 [hep-ph]} \BibitemShut {NoStop}%
\bibitem [{Note2()}]{Note2}%
  \BibitemOpen
  \bibinfo {note} {The vacuum superconductivity at QCD~\cite {Chernodub:2010qx}
  and Electroweak~\cite {Chernodub:2010qx,Chernodub:2012fi} scales is similar
  to reentrant superconductivity which is suggested to occur in clean
  superconducting materials in very high magnetic fields~\cite
  {Rasolt1992}.}\BibitemShut {Stop}%
\bibitem [{\citenamefont {Achucarro}\ \emph {et~al.}(1994)\citenamefont
  {Achucarro}, \citenamefont {Gregory}, \citenamefont {Harvey},\ and\
  \citenamefont {Kuijken}}]{Achucarro:1993bu}%
  \BibitemOpen
  \bibfield  {author} {\bibinfo {author} {\bibfnamefont {A.}~\bibnamefont
  {Achucarro}}, \bibinfo {author} {\bibfnamefont {R.}~\bibnamefont {Gregory}},
  \bibinfo {author} {\bibfnamefont {J.~A.}\ \bibnamefont {Harvey}}, \ and\
  \bibinfo {author} {\bibfnamefont {K.}~\bibnamefont {Kuijken}},\ }\href
  {\doibase 10.1103/PhysRevLett.72.3646} {\bibfield  {journal} {\bibinfo
  {journal} {Phys. Rev. Lett.}\ }\textbf {\bibinfo {volume} {72}},\ \bibinfo
  {pages} {3646} (\bibinfo {year} {1994})},\ \Eprint
  {http://arxiv.org/abs/hep-th/9312034} {arXiv:hep-th/9312034} \BibitemShut
  {NoStop}%
\bibitem [{\citenamefont {Perkins}(1993)}]{Perkins:1993qz}%
  \BibitemOpen
  \bibfield  {author} {\bibinfo {author} {\bibfnamefont {W.~B.}\ \bibnamefont
  {Perkins}},\ }\href {\doibase 10.1103/PhysRevD.47.R5224} {\bibfield
  {journal} {\bibinfo  {journal} {Phys. Rev. D}\ }\textbf {\bibinfo {volume}
  {47}},\ \bibinfo {pages} {R5224} (\bibinfo {year} {1993})}\BibitemShut
  {NoStop}%
\bibitem [{\citenamefont {Olesen}(1993)}]{Olesen:1993ra}%
  \BibitemOpen
  \bibfield  {author} {\bibinfo {author} {\bibfnamefont {P.}~\bibnamefont
  {Olesen}},\ }\href@noop {} {\  (\bibinfo {year} {1993})},\ \Eprint
  {http://arxiv.org/abs/hep-ph/9310275} {arXiv:hep-ph/9310275} \BibitemShut
  {NoStop}%
\bibitem [{\citenamefont {Garaud}\ and\ \citenamefont
  {Volkov}(2010)}]{Garaud:2009uy}%
  \BibitemOpen
  \bibfield  {author} {\bibinfo {author} {\bibfnamefont {J.}~\bibnamefont
  {Garaud}}\ and\ \bibinfo {author} {\bibfnamefont {M.~S.}\ \bibnamefont
  {Volkov}},\ }\href {\doibase 10.1016/j.nuclphysb.2009.10.003} {\bibfield
  {journal} {\bibinfo  {journal} {Nucl. Phys. B}\ }\textbf {\bibinfo {volume}
  {826}},\ \bibinfo {pages} {174} (\bibinfo {year} {2010})},\ \Eprint
  {http://arxiv.org/abs/0906.2996} {arXiv:0906.2996 [hep-th]} \BibitemShut
  {NoStop}%
\bibitem [{\citenamefont {Salam}\ and\ \citenamefont
  {Strathdee}(1975)}]{Salam:1974xe}%
  \BibitemOpen
  \bibfield  {author} {\bibinfo {author} {\bibfnamefont {A.}~\bibnamefont
  {Salam}}\ and\ \bibinfo {author} {\bibfnamefont {J.~A.}\ \bibnamefont
  {Strathdee}},\ }\href {\doibase 10.1016/0550-3213(75)90642-2} {\bibfield
  {journal} {\bibinfo  {journal} {Nucl. Phys. B}\ }\textbf {\bibinfo {volume}
  {90}},\ \bibinfo {pages} {203} (\bibinfo {year} {1975})}\BibitemShut
  {NoStop}%
\bibitem [{\citenamefont {Linde}(1976)}]{Linde:1975gx}%
  \BibitemOpen
  \bibfield  {author} {\bibinfo {author} {\bibfnamefont {A.~D.}\ \bibnamefont
  {Linde}},\ }\href {\doibase 10.1016/0370-2693(76)90678-X} {\bibfield
  {journal} {\bibinfo  {journal} {Phys. Lett. B}\ }\textbf {\bibinfo {volume}
  {62}},\ \bibinfo {pages} {435} (\bibinfo {year} {1976})}\BibitemShut
  {NoStop}%
\bibitem [{\citenamefont {Olesen}(1991)}]{Olesen:1991df}%
  \BibitemOpen
  \bibfield  {author} {\bibinfo {author} {\bibfnamefont {P.}~\bibnamefont
  {Olesen}},\ }\href {\doibase 10.1016/0370-2693(91)91595-M} {\bibfield
  {journal} {\bibinfo  {journal} {Phys. Lett. B}\ }\textbf {\bibinfo {volume}
  {268}},\ \bibinfo {pages} {389} (\bibinfo {year} {1991})}\BibitemShut
  {NoStop}%
\bibitem [{\citenamefont {Van~Doorsselaere}(2013)}]{VanDoorsselaere:2012zb}%
  \BibitemOpen
  \bibfield  {author} {\bibinfo {author} {\bibfnamefont {J.}~\bibnamefont
  {Van~Doorsselaere}},\ }\href {\doibase 10.1103/PhysRevD.88.025013} {\bibfield
   {journal} {\bibinfo  {journal} {Phys. Rev. D}\ }\textbf {\bibinfo {volume}
  {88}},\ \bibinfo {pages} {025013} (\bibinfo {year} {2013})},\ \Eprint
  {http://arxiv.org/abs/1206.6205} {arXiv:1206.6205 [hep-ph]} \BibitemShut
  {NoStop}%
\bibitem [{\citenamefont {Vachaspati}(1991)}]{Vachaspati:1991nm}%
  \BibitemOpen
  \bibfield  {author} {\bibinfo {author} {\bibfnamefont {T.}~\bibnamefont
  {Vachaspati}},\ }\href {\doibase 10.1016/0370-2693(91)90051-Q} {\bibfield
  {journal} {\bibinfo  {journal} {Phys. Lett. B}\ }\textbf {\bibinfo {volume}
  {265}},\ \bibinfo {pages} {258} (\bibinfo {year} {1991})}\BibitemShut
  {NoStop}%
\bibitem [{\citenamefont {Grasso}\ and\ \citenamefont
  {Rubinstein}(2001)}]{Grasso:2000wj}%
  \BibitemOpen
  \bibfield  {author} {\bibinfo {author} {\bibfnamefont {D.}~\bibnamefont
  {Grasso}}\ and\ \bibinfo {author} {\bibfnamefont {H.~R.}\ \bibnamefont
  {Rubinstein}},\ }\href {\doibase 10.1016/S0370-1573(00)00110-1} {\bibfield
  {journal} {\bibinfo  {journal} {Phys. Rept.}\ }\textbf {\bibinfo {volume}
  {348}},\ \bibinfo {pages} {163} (\bibinfo {year} {2001})},\ \Eprint
  {http://arxiv.org/abs/astro-ph/0009061} {arXiv:astro-ph/0009061} \BibitemShut
  {NoStop}%
\bibitem [{\citenamefont {Maldacena}(2021)}]{Maldacena:2020skw}%
  \BibitemOpen
  \bibfield  {author} {\bibinfo {author} {\bibfnamefont {J.}~\bibnamefont
  {Maldacena}},\ }\href {\doibase 10.1007/JHEP04(2021)079} {\bibfield
  {journal} {\bibinfo  {journal} {JHEP}\ }\textbf {\bibinfo {volume} {04}},\
  \bibinfo {pages} {079} (\bibinfo {year} {2021})},\ \Eprint
  {http://arxiv.org/abs/2004.06084} {arXiv:2004.06084 [hep-th]} \BibitemShut
  {NoStop}%
\bibitem [{\citenamefont {Ghosh}\ \emph {et~al.}(2021)\citenamefont {Ghosh},
  \citenamefont {Thalapillil},\ and\ \citenamefont {Ullah}}]{Ghosh:2020tdu}%
  \BibitemOpen
  \bibfield  {author} {\bibinfo {author} {\bibfnamefont {D.}~\bibnamefont
  {Ghosh}}, \bibinfo {author} {\bibfnamefont {A.}~\bibnamefont {Thalapillil}},
  \ and\ \bibinfo {author} {\bibfnamefont {F.}~\bibnamefont {Ullah}},\ }\href
  {\doibase 10.1103/PhysRevD.103.023006} {\bibfield  {journal} {\bibinfo
  {journal} {Phys. Rev. D}\ }\textbf {\bibinfo {volume} {103}},\ \bibinfo
  {pages} {023006} (\bibinfo {year} {2021})},\ \Eprint
  {http://arxiv.org/abs/2009.03363} {arXiv:2009.03363 [hep-ph]} \BibitemShut
  {NoStop}%
\bibitem [{\citenamefont {Ho}\ and\ \citenamefont
  {Rajantie}(2020)}]{Ho:2020ltr}%
  \BibitemOpen
  \bibfield  {author} {\bibinfo {author} {\bibfnamefont {D.~L.~J.}\
  \bibnamefont {Ho}}\ and\ \bibinfo {author} {\bibfnamefont {A.}~\bibnamefont
  {Rajantie}},\ }\href {\doibase 10.1103/PhysRevD.102.053002} {\bibfield
  {journal} {\bibinfo  {journal} {Phys. Rev. D}\ }\textbf {\bibinfo {volume}
  {102}},\ \bibinfo {pages} {053002} (\bibinfo {year} {2020})},\ \Eprint
  {http://arxiv.org/abs/2005.03125} {arXiv:2005.03125 [hep-th]} \BibitemShut
  {NoStop}%
\bibitem [{\citenamefont {Skalozub}(2014)}]{Skalozub:2014epa}%
  \BibitemOpen
  \bibfield  {author} {\bibinfo {author} {\bibfnamefont {V.~V.}\ \bibnamefont
  {Skalozub}},\ }\href {\doibase 10.1134/S1063778814070151} {\bibfield
  {journal} {\bibinfo  {journal} {Phys. Atom. Nucl.}\ }\textbf {\bibinfo
  {volume} {77}},\ \bibinfo {pages} {901} (\bibinfo {year} {2014})}\BibitemShut
  {NoStop}%
\bibitem [{\citenamefont {Kajantie}\ \emph {et~al.}(1999)\citenamefont
  {Kajantie}, \citenamefont {Laine}, \citenamefont {Peisa}, \citenamefont
  {Rummukainen},\ and\ \citenamefont {Shaposhnikov}}]{Kajantie:1998rz}%
  \BibitemOpen
  \bibfield  {author} {\bibinfo {author} {\bibfnamefont {K.}~\bibnamefont
  {Kajantie}}, \bibinfo {author} {\bibfnamefont {M.}~\bibnamefont {Laine}},
  \bibinfo {author} {\bibfnamefont {J.}~\bibnamefont {Peisa}}, \bibinfo
  {author} {\bibfnamefont {K.}~\bibnamefont {Rummukainen}}, \ and\ \bibinfo
  {author} {\bibfnamefont {M.~E.}\ \bibnamefont {Shaposhnikov}},\ }\href
  {\doibase 10.1016/S0550-3213(98)00854-2} {\bibfield  {journal} {\bibinfo
  {journal} {Nucl. Phys. B}\ }\textbf {\bibinfo {volume} {544}},\ \bibinfo
  {pages} {357} (\bibinfo {year} {1999})}\BibitemShut {NoStop}%
\bibitem [{\citenamefont {Hidaka}\ and\ \citenamefont
  {Yamamoto}(2013)}]{Hidaka:2012mz}%
  \BibitemOpen
  \bibfield  {author} {\bibinfo {author} {\bibfnamefont {Y.}~\bibnamefont
  {Hidaka}}\ and\ \bibinfo {author} {\bibfnamefont {A.}~\bibnamefont
  {Yamamoto}},\ }\href {\doibase 10.1103/PhysRevD.87.094502} {\bibfield
  {journal} {\bibinfo  {journal} {Phys. Rev. D}\ }\textbf {\bibinfo {volume}
  {87}},\ \bibinfo {pages} {094502} (\bibinfo {year} {2013})},\ \Eprint
  {http://arxiv.org/abs/1209.0007} {arXiv:1209.0007 [hep-ph]} \BibitemShut
  {NoStop}%
\bibitem [{\citenamefont {Andreichikov}\ \emph {et~al.}(2013)\citenamefont
  {Andreichikov}, \citenamefont {Kerbikov}, \citenamefont {Orlovsky},\ and\
  \citenamefont {Simonov}}]{Andreichikov:2013zba}%
  \BibitemOpen
  \bibfield  {author} {\bibinfo {author} {\bibfnamefont {M.~A.}\ \bibnamefont
  {Andreichikov}}, \bibinfo {author} {\bibfnamefont {B.~O.}\ \bibnamefont
  {Kerbikov}}, \bibinfo {author} {\bibfnamefont {V.~D.}\ \bibnamefont
  {Orlovsky}}, \ and\ \bibinfo {author} {\bibfnamefont {Y.~A.}\ \bibnamefont
  {Simonov}},\ }\href {\doibase 10.1103/PhysRevD.87.094029} {\bibfield
  {journal} {\bibinfo  {journal} {Phys. Rev. D}\ }\textbf {\bibinfo {volume}
  {87}},\ \bibinfo {pages} {094029} (\bibinfo {year} {2013})},\ \Eprint
  {http://arxiv.org/abs/1304.2533} {arXiv:1304.2533 [hep-ph]} \BibitemShut
  {NoStop}%
\bibitem [{\citenamefont {Bali}\ \emph {et~al.}(2018)\citenamefont {Bali},
  \citenamefont {Brandt}, \citenamefont {Endr\H{o}di},\ and\ \citenamefont
  {Gl\"a\ss{}le}}]{Bali:2017ian}%
  \BibitemOpen
  \bibfield  {author} {\bibinfo {author} {\bibfnamefont {G.~S.}\ \bibnamefont
  {Bali}}, \bibinfo {author} {\bibfnamefont {B.~B.}\ \bibnamefont {Brandt}},
  \bibinfo {author} {\bibfnamefont {G.}~\bibnamefont {Endr\H{o}di}}, \ and\
  \bibinfo {author} {\bibfnamefont {B.}~\bibnamefont {Gl\"a\ss{}le}},\ }\href
  {\doibase 10.1103/PhysRevD.97.034505} {\bibfield  {journal} {\bibinfo
  {journal} {Phys. Rev. D}\ }\textbf {\bibinfo {volume} {97}},\ \bibinfo
  {pages} {034505} (\bibinfo {year} {2018})}\BibitemShut {NoStop}%
\bibitem [{\citenamefont {Chernodub}(2014)}]{Chernodub:2013uja}%
  \BibitemOpen
  \bibfield  {author} {\bibinfo {author} {\bibfnamefont {M.~N.}\ \bibnamefont
  {Chernodub}},\ }\href {\doibase 10.1103/PhysRevD.89.018501} {\bibfield
  {journal} {\bibinfo  {journal} {Phys. Rev. D}\ }\textbf {\bibinfo {volume}
  {89}},\ \bibinfo {pages} {018501} (\bibinfo {year} {2014})},\ \Eprint
  {http://arxiv.org/abs/1309.4071} {arXiv:1309.4071 [hep-ph]} \BibitemShut
  {NoStop}%
\bibitem [{\citenamefont {Tiesinga}\ \emph {et~al.}(2021)\citenamefont
  {Tiesinga}, \citenamefont {Mohr}, \citenamefont {Newell},\ and\ \citenamefont
  {Taylor}}]{CODATA2018}%
  \BibitemOpen
  \bibfield  {author} {\bibinfo {author} {\bibfnamefont {E.}~\bibnamefont
  {Tiesinga}}, \bibinfo {author} {\bibfnamefont {P.~J.}\ \bibnamefont {Mohr}},
  \bibinfo {author} {\bibfnamefont {D.~B.}\ \bibnamefont {Newell}}, \ and\
  \bibinfo {author} {\bibfnamefont {B.~N.}\ \bibnamefont {Taylor}},\
  }\href@noop {} {\bibfield  {journal} {\bibinfo  {journal} {Journal of
  Physical and Chemical Reference Data}\ }\textbf {\bibinfo {volume} {50}},\
  \bibinfo {pages} {033105} (\bibinfo {year} {2021})}\BibitemShut {NoStop}%
\bibitem [{\citenamefont {Bunk}\ \emph {et~al.}(1992)\citenamefont {Bunk},
  \citenamefont {Ilgenfritz}, \citenamefont {Kripfganz},\ and\ \citenamefont
  {Schiller}}]{Bunk:1992xt}%
  \BibitemOpen
  \bibfield  {author} {\bibinfo {author} {\bibfnamefont {B.}~\bibnamefont
  {Bunk}}, \bibinfo {author} {\bibfnamefont {E.-M.}\ \bibnamefont
  {Ilgenfritz}}, \bibinfo {author} {\bibfnamefont {J.}~\bibnamefont
  {Kripfganz}}, \ and\ \bibinfo {author} {\bibfnamefont {A.}~\bibnamefont
  {Schiller}},\ }\href {\doibase 10.1016/0370-2693(92)90447-C} {\bibfield
  {journal} {\bibinfo  {journal} {Phys. Lett. B}\ }\textbf {\bibinfo {volume}
  {284}},\ \bibinfo {pages} {371} (\bibinfo {year} {1992})}\BibitemShut
  {NoStop}%
\bibitem [{\citenamefont {Bunk}\ \emph {et~al.}(1993)\citenamefont {Bunk},
  \citenamefont {Ilgenfritz}, \citenamefont {Kripfganz},\ and\ \citenamefont
  {Schiller}}]{Bunk:1992kf}%
  \BibitemOpen
  \bibfield  {author} {\bibinfo {author} {\bibfnamefont {B.}~\bibnamefont
  {Bunk}}, \bibinfo {author} {\bibfnamefont {E.-M.}\ \bibnamefont
  {Ilgenfritz}}, \bibinfo {author} {\bibfnamefont {J.}~\bibnamefont
  {Kripfganz}}, \ and\ \bibinfo {author} {\bibfnamefont {A.}~\bibnamefont
  {Schiller}},\ }\href {\doibase 10.1016/0550-3213(93)90043-O} {\bibfield
  {journal} {\bibinfo  {journal} {Nucl. Phys. B}\ }\textbf {\bibinfo {volume}
  {403}},\ \bibinfo {pages} {453} (\bibinfo {year} {1993})}\BibitemShut
  {NoStop}%
\bibitem [{\citenamefont {Fodor}\ \emph {et~al.}(1995)\citenamefont {Fodor},
  \citenamefont {Hein}, \citenamefont {Jansen}, \citenamefont {Jaster},\ and\
  \citenamefont {Montvay}}]{Fodor:1994sj}%
  \BibitemOpen
  \bibfield  {author} {\bibinfo {author} {\bibfnamefont {Z.}~\bibnamefont
  {Fodor}}, \bibinfo {author} {\bibfnamefont {J.}~\bibnamefont {Hein}},
  \bibinfo {author} {\bibfnamefont {K.}~\bibnamefont {Jansen}}, \bibinfo
  {author} {\bibfnamefont {A.}~\bibnamefont {Jaster}}, \ and\ \bibinfo {author}
  {\bibfnamefont {I.}~\bibnamefont {Montvay}},\ }\href {\doibase
  10.1016/0550-3213(95)00038-T} {\bibfield  {journal} {\bibinfo  {journal}
  {Nucl. Phys. B}\ }\textbf {\bibinfo {volume} {439}},\ \bibinfo {pages} {147}
  (\bibinfo {year} {1995})},\ \Eprint {http://arxiv.org/abs/hep-lat/9409017}
  {arXiv:hep-lat/9409017} \BibitemShut {NoStop}%
\bibitem [{\citenamefont {Fodor}\ \emph {et~al.}(1994)\citenamefont {Fodor},
  \citenamefont {Hein}, \citenamefont {Jansen}, \citenamefont {Jaster},
  \citenamefont {Montvay},\ and\ \citenamefont {Csikor}}]{Fodor:1994dm}%
  \BibitemOpen
  \bibfield  {author} {\bibinfo {author} {\bibfnamefont {Z.}~\bibnamefont
  {Fodor}}, \bibinfo {author} {\bibfnamefont {J.}~\bibnamefont {Hein}},
  \bibinfo {author} {\bibfnamefont {K.}~\bibnamefont {Jansen}}, \bibinfo
  {author} {\bibfnamefont {A.}~\bibnamefont {Jaster}}, \bibinfo {author}
  {\bibfnamefont {I.}~\bibnamefont {Montvay}}, \ and\ \bibinfo {author}
  {\bibfnamefont {F.}~\bibnamefont {Csikor}},\ }\href {\doibase
  10.1016/0370-2693(94)90706-4} {\bibfield  {journal} {\bibinfo  {journal}
  {Phys. Lett. B}\ }\textbf {\bibinfo {volume} {334}},\ \bibinfo {pages} {405}
  (\bibinfo {year} {1994})},\ \Eprint {http://arxiv.org/abs/hep-lat/9405021}
  {arXiv:hep-lat/9405021} \BibitemShut {NoStop}%
\bibitem [{\citenamefont {Csikor}\ \emph {et~al.}(1996)\citenamefont {Csikor},
  \citenamefont {Fodor}, \citenamefont {Hein}, \citenamefont {Jaster},\ and\
  \citenamefont {Montvay}}]{Csikor:1996sp}%
  \BibitemOpen
  \bibfield  {author} {\bibinfo {author} {\bibfnamefont {F.}~\bibnamefont
  {Csikor}}, \bibinfo {author} {\bibfnamefont {Z.}~\bibnamefont {Fodor}},
  \bibinfo {author} {\bibfnamefont {J.}~\bibnamefont {Hein}}, \bibinfo {author}
  {\bibfnamefont {A.}~\bibnamefont {Jaster}}, \ and\ \bibinfo {author}
  {\bibfnamefont {I.}~\bibnamefont {Montvay}},\ }\href {\doibase
  10.1016/0550-3213(96)00285-4} {\bibfield  {journal} {\bibinfo  {journal}
  {Nucl. Phys. B}\ }\textbf {\bibinfo {volume} {474}},\ \bibinfo {pages} {421}
  (\bibinfo {year} {1996})},\ \Eprint {http://arxiv.org/abs/hep-lat/9601016}
  {arXiv:hep-lat/9601016} \BibitemShut {NoStop}%
\bibitem [{\citenamefont {Aoki}\ \emph {et~al.}(1999)\citenamefont {Aoki},
  \citenamefont {Csikor}, \citenamefont {Fodor},\ and\ \citenamefont
  {Ukawa}}]{Aoki:1999fi}%
  \BibitemOpen
  \bibfield  {author} {\bibinfo {author} {\bibfnamefont {Y.}~\bibnamefont
  {Aoki}}, \bibinfo {author} {\bibfnamefont {F.}~\bibnamefont {Csikor}},
  \bibinfo {author} {\bibfnamefont {Z.}~\bibnamefont {Fodor}}, \ and\ \bibinfo
  {author} {\bibfnamefont {A.}~\bibnamefont {Ukawa}},\ }\href {\doibase
  10.1103/PhysRevD.60.013001} {\bibfield  {journal} {\bibinfo  {journal} {Phys.
  Rev. D}\ }\textbf {\bibinfo {volume} {60}},\ \bibinfo {pages} {013001}
  (\bibinfo {year} {1999})},\ \Eprint {http://arxiv.org/abs/hep-lat/9901021}
  {arXiv:hep-lat/9901021} \BibitemShut {NoStop}%
\bibitem [{\citenamefont {Sakharov}(1967)}]{Sakharov:1967dj}%
  \BibitemOpen
  \bibfield  {author} {\bibinfo {author} {\bibfnamefont {A.~D.}\ \bibnamefont
  {Sakharov}},\ }\href {\doibase 10.1070/PU1991v034n05ABEH002497} {\bibfield
  {journal} {\bibinfo  {journal} {Pisma Zh. Eksp. Teor. Fiz.}\ }\textbf
  {\bibinfo {volume} {5}},\ \bibinfo {pages} {32} (\bibinfo {year}
  {1967})}\BibitemShut {NoStop}%
\bibitem [{\citenamefont {Rubakov}\ and\ \citenamefont
  {Shaposhnikov}(1996)}]{Rubakov:1996vz}%
  \BibitemOpen
  \bibfield  {author} {\bibinfo {author} {\bibfnamefont {V.~A.}\ \bibnamefont
  {Rubakov}}\ and\ \bibinfo {author} {\bibfnamefont {M.~E.}\ \bibnamefont
  {Shaposhnikov}},\ }\href {\doibase 10.1070/PU1996v039n05ABEH000145}
  {\bibfield  {journal} {\bibinfo  {journal} {Usp. Fiz. Nauk}\ }\textbf
  {\bibinfo {volume} {166}},\ \bibinfo {pages} {493} (\bibinfo {year}
  {1996})},\ \Eprint {http://arxiv.org/abs/hep-ph/9603208}
  {arXiv:hep-ph/9603208} \BibitemShut {NoStop}%
\bibitem [{\citenamefont {Kajantie}\ \emph {et~al.}(1993)\citenamefont
  {Kajantie}, \citenamefont {Rummukainen},\ and\ \citenamefont
  {Shaposhnikov}}]{Kajantie:1993ag}%
  \BibitemOpen
  \bibfield  {author} {\bibinfo {author} {\bibfnamefont {K.}~\bibnamefont
  {Kajantie}}, \bibinfo {author} {\bibfnamefont {K.}~\bibnamefont
  {Rummukainen}}, \ and\ \bibinfo {author} {\bibfnamefont {M.~E.}\ \bibnamefont
  {Shaposhnikov}},\ }\href {\doibase 10.1016/0550-3213(93)90062-T} {\bibfield
  {journal} {\bibinfo  {journal} {Nucl. Phys. B}\ }\textbf {\bibinfo {volume}
  {407}},\ \bibinfo {pages} {356} (\bibinfo {year} {1993})},\ \Eprint
  {http://arxiv.org/abs/hep-ph/9305345} {arXiv:hep-ph/9305345} \BibitemShut
  {NoStop}%
\bibitem [{\citenamefont {Ilgenfritz}\ \emph {et~al.}(1995)\citenamefont
  {Ilgenfritz}, \citenamefont {Kripfganz}, \citenamefont {Perlt},\ and\
  \citenamefont {Schiller}}]{Ilgenfritz:1995sh}%
  \BibitemOpen
  \bibfield  {author} {\bibinfo {author} {\bibfnamefont {E.-M.}\ \bibnamefont
  {Ilgenfritz}}, \bibinfo {author} {\bibfnamefont {J.}~\bibnamefont
  {Kripfganz}}, \bibinfo {author} {\bibfnamefont {H.}~\bibnamefont {Perlt}}, \
  and\ \bibinfo {author} {\bibfnamefont {A.}~\bibnamefont {Schiller}},\ }\href
  {\doibase 10.1016/0370-2693(95)00856-G} {\bibfield  {journal} {\bibinfo
  {journal} {Phys. Lett. B}\ }\textbf {\bibinfo {volume} {356}},\ \bibinfo
  {pages} {561} (\bibinfo {year} {1995})},\ \Eprint
  {http://arxiv.org/abs/hep-lat/9506023} {arXiv:hep-lat/9506023} \BibitemShut
  {NoStop}%
\bibitem [{\citenamefont {Gurtler}\ \emph {et~al.}(1997)\citenamefont
  {Gurtler}, \citenamefont {Ilgenfritz}, \citenamefont {Kripfganz},
  \citenamefont {Perlt},\ and\ \citenamefont {Schiller}}]{Gurtler:1996wx}%
  \BibitemOpen
  \bibfield  {author} {\bibinfo {author} {\bibfnamefont {M.}~\bibnamefont
  {Gurtler}}, \bibinfo {author} {\bibfnamefont {E.-M.}\ \bibnamefont
  {Ilgenfritz}}, \bibinfo {author} {\bibfnamefont {J.}~\bibnamefont
  {Kripfganz}}, \bibinfo {author} {\bibfnamefont {H.}~\bibnamefont {Perlt}}, \
  and\ \bibinfo {author} {\bibfnamefont {A.}~\bibnamefont {Schiller}},\ }\href
  {\doibase 10.1016/S0550-3213(96)00594-9} {\bibfield  {journal} {\bibinfo
  {journal} {Nucl. Phys. B}\ }\textbf {\bibinfo {volume} {483}},\ \bibinfo
  {pages} {383} (\bibinfo {year} {1997})},\ \Eprint
  {http://arxiv.org/abs/hep-lat/9605042} {arXiv:hep-lat/9605042} \BibitemShut
  {NoStop}%
\bibitem [{\citenamefont {Kajantie}\ \emph {et~al.}(1997)\citenamefont
  {Kajantie}, \citenamefont {Laine}, \citenamefont {Rummukainen},\ and\
  \citenamefont {Shaposhnikov}}]{Kajantie:1996qd}%
  \BibitemOpen
  \bibfield  {author} {\bibinfo {author} {\bibfnamefont {K.}~\bibnamefont
  {Kajantie}}, \bibinfo {author} {\bibfnamefont {M.}~\bibnamefont {Laine}},
  \bibinfo {author} {\bibfnamefont {K.}~\bibnamefont {Rummukainen}}, \ and\
  \bibinfo {author} {\bibfnamefont {M.~E.}\ \bibnamefont {Shaposhnikov}},\
  }\href {\doibase 10.1016/S0550-3213(97)00164-8} {\bibfield  {journal}
  {\bibinfo  {journal} {Nucl. Phys. B}\ }\textbf {\bibinfo {volume} {493}},\
  \bibinfo {pages} {413} (\bibinfo {year} {1997})},\ \Eprint
  {http://arxiv.org/abs/hep-lat/9612006} {arXiv:hep-lat/9612006} \BibitemShut
  {NoStop}%
\bibitem [{\citenamefont {Kajantie}\ \emph {et~al.}(1996)\citenamefont
  {Kajantie}, \citenamefont {Laine}, \citenamefont {Rummukainen},\ and\
  \citenamefont {Shaposhnikov}}]{Kajantie:1996mn}%
  \BibitemOpen
  \bibfield  {author} {\bibinfo {author} {\bibfnamefont {K.}~\bibnamefont
  {Kajantie}}, \bibinfo {author} {\bibfnamefont {M.}~\bibnamefont {Laine}},
  \bibinfo {author} {\bibfnamefont {K.}~\bibnamefont {Rummukainen}}, \ and\
  \bibinfo {author} {\bibfnamefont {M.~E.}\ \bibnamefont {Shaposhnikov}},\
  }\href {\doibase 10.1103/PhysRevLett.77.2887} {\bibfield  {journal} {\bibinfo
   {journal} {Phys. Rev. Lett.}\ }\textbf {\bibinfo {volume} {77}},\ \bibinfo
  {pages} {2887} (\bibinfo {year} {1996})},\ \Eprint
  {http://arxiv.org/abs/hep-ph/9605288} {arXiv:hep-ph/9605288} \BibitemShut
  {NoStop}%
\bibitem [{\citenamefont {D'Onofrio}\ and\ \citenamefont
  {Rummukainen}(2016)}]{DOnofrio:2015gop}%
  \BibitemOpen
  \bibfield  {author} {\bibinfo {author} {\bibfnamefont {M.}~\bibnamefont
  {D'Onofrio}}\ and\ \bibinfo {author} {\bibfnamefont {K.}~\bibnamefont
  {Rummukainen}},\ }\href {\doibase 10.1103/PhysRevD.93.025003} {\bibfield
  {journal} {\bibinfo  {journal} {Phys. Rev. D}\ }\textbf {\bibinfo {volume}
  {93}},\ \bibinfo {pages} {025003} (\bibinfo {year} {2016})},\ \Eprint
  {http://arxiv.org/abs/1508.07161} {arXiv:1508.07161 [hep-ph]} \BibitemShut
  {NoStop}%
\bibitem [{\citenamefont {Langguth}\ \emph {et~al.}(1986)\citenamefont
  {Langguth}, \citenamefont {Montvay},\ and\ \citenamefont
  {Weisz}}]{Langguth:1985dr}%
  \BibitemOpen
  \bibfield  {author} {\bibinfo {author} {\bibfnamefont {W.}~\bibnamefont
  {Langguth}}, \bibinfo {author} {\bibfnamefont {I.}~\bibnamefont {Montvay}}, \
  and\ \bibinfo {author} {\bibfnamefont {P.}~\bibnamefont {Weisz}},\ }\href
  {\doibase 10.1016/0550-3213(86)90430-X} {\bibfield  {journal} {\bibinfo
  {journal} {Nucl. Phys. B}\ }\textbf {\bibinfo {volume} {277}},\ \bibinfo
  {pages} {11} (\bibinfo {year} {1986})}\BibitemShut {NoStop}%
\bibitem [{\citenamefont {Bali}\ \emph {et~al.}(2012)\citenamefont {Bali},
  \citenamefont {Bruckmann}, \citenamefont {Endrodi}, \citenamefont {Fodor},
  \citenamefont {Katz}, \citenamefont {Krieg}, \citenamefont {Schafer},\ and\
  \citenamefont {Szabo}}]{Bali:2011qj}%
  \BibitemOpen
  \bibfield  {author} {\bibinfo {author} {\bibfnamefont {G.~S.}\ \bibnamefont
  {Bali}}, \bibinfo {author} {\bibfnamefont {F.}~\bibnamefont {Bruckmann}},
  \bibinfo {author} {\bibfnamefont {G.}~\bibnamefont {Endrodi}}, \bibinfo
  {author} {\bibfnamefont {Z.}~\bibnamefont {Fodor}}, \bibinfo {author}
  {\bibfnamefont {S.~D.}\ \bibnamefont {Katz}}, \bibinfo {author}
  {\bibfnamefont {S.}~\bibnamefont {Krieg}}, \bibinfo {author} {\bibfnamefont
  {A.}~\bibnamefont {Schafer}}, \ and\ \bibinfo {author} {\bibfnamefont
  {K.~K.}\ \bibnamefont {Szabo}},\ }\href {\doibase 10.1007/JHEP02(2012)044}
  {\bibfield  {journal} {\bibinfo  {journal} {JHEP}\ }\textbf {\bibinfo
  {volume} {02}},\ \bibinfo {pages} {044} (\bibinfo {year} {2012})},\ \Eprint
  {http://arxiv.org/abs/1111.4956} {arXiv:1111.4956 [hep-lat]} \BibitemShut
  {NoStop}%
\bibitem [{\citenamefont {Gattringer}\ and\ \citenamefont
  {Lang}(2010)}]{Gattringer:2010zz}%
  \BibitemOpen
  \bibfield  {author} {\bibinfo {author} {\bibfnamefont {C.}~\bibnamefont
  {Gattringer}}\ and\ \bibinfo {author} {\bibfnamefont {C.~B.}\ \bibnamefont
  {Lang}},\ }\href {\doibase 10.1007/978-3-642-01850-3} {\emph {\bibinfo
  {title} {{Quantum chromodynamics on the lattice}}}},\ Vol.\ \bibinfo {volume}
  {788}\ (\bibinfo  {publisher} {Springer},\ \bibinfo {address} {Berlin},\
  \bibinfo {year} {2010})\BibitemShut {NoStop}%
\bibitem [{\citenamefont {Frohlich}\ \emph {et~al.}(1980)\citenamefont
  {Frohlich}, \citenamefont {Morchio},\ and\ \citenamefont
  {Strocchi}}]{Frohlich:1980gj}%
  \BibitemOpen
  \bibfield  {author} {\bibinfo {author} {\bibfnamefont {J.}~\bibnamefont
  {Frohlich}}, \bibinfo {author} {\bibfnamefont {G.}~\bibnamefont {Morchio}}, \
  and\ \bibinfo {author} {\bibfnamefont {F.}~\bibnamefont {Strocchi}},\ }\href
  {\doibase 10.1016/0370-2693(80)90594-8} {\bibfield  {journal} {\bibinfo
  {journal} {Phys. Lett. B}\ }\textbf {\bibinfo {volume} {97}},\ \bibinfo
  {pages} {249} (\bibinfo {year} {1980})}\BibitemShut {NoStop}%
\bibitem [{\citenamefont {Frohlich}\ \emph {et~al.}(1981)\citenamefont
  {Frohlich}, \citenamefont {Morchio},\ and\ \citenamefont
  {Strocchi}}]{Frohlich:1981yi}%
  \BibitemOpen
  \bibfield  {author} {\bibinfo {author} {\bibfnamefont {J.}~\bibnamefont
  {Frohlich}}, \bibinfo {author} {\bibfnamefont {G.}~\bibnamefont {Morchio}}, \
  and\ \bibinfo {author} {\bibfnamefont {F.}~\bibnamefont {Strocchi}},\ }\href
  {\doibase 10.1016/0550-3213(81)90448-X} {\bibfield  {journal} {\bibinfo
  {journal} {Nucl. Phys. B}\ }\textbf {\bibinfo {volume} {190}},\ \bibinfo
  {pages} {553} (\bibinfo {year} {1981})}\BibitemShut {NoStop}%
\bibitem [{\citenamefont {Woloshyn}(2017)}]{Woloshyn:2017rhe}%
  \BibitemOpen
  \bibfield  {author} {\bibinfo {author} {\bibfnamefont {R.~M.}\ \bibnamefont
  {Woloshyn}},\ }\href {\doibase 10.1103/PhysRevD.95.054507} {\bibfield
  {journal} {\bibinfo  {journal} {Phys. Rev. D}\ }\textbf {\bibinfo {volume}
  {95}},\ \bibinfo {pages} {054507} (\bibinfo {year} {2017})},\ \Eprint
  {http://arxiv.org/abs/1702.01693} {arXiv:1702.01693 [hep-lat]} \BibitemShut
  {NoStop}%
\bibitem [{\citenamefont {Lewis}\ and\ \citenamefont
  {Woloshyn}(2018)}]{Lewis:2018srt}%
  \BibitemOpen
  \bibfield  {author} {\bibinfo {author} {\bibfnamefont {R.}~\bibnamefont
  {Lewis}}\ and\ \bibinfo {author} {\bibfnamefont {R.~M.}\ \bibnamefont
  {Woloshyn}},\ }\href {\doibase 10.1103/PhysRevD.98.034502} {\bibfield
  {journal} {\bibinfo  {journal} {Phys. Rev. D}\ }\textbf {\bibinfo {volume}
  {98}},\ \bibinfo {pages} {034502} (\bibinfo {year} {2018})},\ \Eprint
  {http://arxiv.org/abs/1806.11380} {arXiv:1806.11380 [hep-lat]} \BibitemShut
  {NoStop}%
\bibitem [{\citenamefont {Maas}(2019)}]{Maas:2017wzi}%
  \BibitemOpen
  \bibfield  {author} {\bibinfo {author} {\bibfnamefont {A.}~\bibnamefont
  {Maas}},\ }\href {\doibase 10.1016/j.ppnp.2019.02.003} {\bibfield  {journal}
  {\bibinfo  {journal} {Prog. Part. Nucl. Phys.}\ }\textbf {\bibinfo {volume}
  {106}},\ \bibinfo {pages} {132} (\bibinfo {year} {2019})},\ \Eprint
  {http://arxiv.org/abs/1712.04721} {arXiv:1712.04721 [hep-ph]} \BibitemShut
  {NoStop}%
\bibitem [{\citenamefont {Durr}(2005)}]{Durr:2004xu}%
  \BibitemOpen
  \bibfield  {author} {\bibinfo {author} {\bibfnamefont {S.}~\bibnamefont
  {Durr}},\ }\href {\doibase 10.1016/j.cpc.2005.06.011} {\bibfield  {journal}
  {\bibinfo  {journal} {Comput. Phys. Commun.}\ }\textbf {\bibinfo {volume}
  {172}},\ \bibinfo {pages} {163} (\bibinfo {year} {2005})}\BibitemShut
  {NoStop}%
\bibitem [{\citenamefont {Workman}\ \emph {et~al.}(2022)\citenamefont {Workman}
  \emph {et~al.}}]{PDG}%
  \BibitemOpen
  \bibfield  {author} {\bibinfo {author} {\bibfnamefont {R.}~\bibnamefont
  {Workman}} \emph {et~al.} (\bibinfo {collaboration} {Particle Data Group}),\
  }\href@noop {} {\bibfield  {journal} {\bibinfo  {journal} {Prog. Theor. Exp.
  Phys.}\ }\textbf {\bibinfo {volume} {083C01}} (\bibinfo {year}
  {2022})}\BibitemShut {NoStop}%
\bibitem [{\citenamefont {Csikor}\ \emph {et~al.}(1999)\citenamefont {Csikor},
  \citenamefont {Fodor},\ and\ \citenamefont {Heitger}}]{Csikor:1998eu}%
  \BibitemOpen
  \bibfield  {author} {\bibinfo {author} {\bibfnamefont {F.}~\bibnamefont
  {Csikor}}, \bibinfo {author} {\bibfnamefont {Z.}~\bibnamefont {Fodor}}, \
  and\ \bibinfo {author} {\bibfnamefont {J.}~\bibnamefont {Heitger}},\ }\href
  {\doibase 10.1103/PhysRevLett.82.21} {\bibfield  {journal} {\bibinfo
  {journal} {Phys. Rev. Lett.}\ }\textbf {\bibinfo {volume} {82}},\ \bibinfo
  {pages} {21} (\bibinfo {year} {1999})},\ \Eprint
  {http://arxiv.org/abs/hep-ph/9809291} {arXiv:hep-ph/9809291} \BibitemShut
  {NoStop}%
\bibitem [{\citenamefont {Osterwalder}\ and\ \citenamefont
  {Seiler}(1978)}]{Osterwalder:1977pc}%
  \BibitemOpen
  \bibfield  {author} {\bibinfo {author} {\bibfnamefont {K.}~\bibnamefont
  {Osterwalder}}\ and\ \bibinfo {author} {\bibfnamefont {E.}~\bibnamefont
  {Seiler}},\ }\href {\doibase 10.1016/0003-4916(78)90039-8} {\bibfield
  {journal} {\bibinfo  {journal} {Annals Phys.}\ }\textbf {\bibinfo {volume}
  {110}},\ \bibinfo {pages} {440} (\bibinfo {year} {1978})}\BibitemShut
  {NoStop}%
\bibitem [{\citenamefont {Fradkin}\ and\ \citenamefont
  {Shenker}(1979)}]{Fradkin:1978dv}%
  \BibitemOpen
  \bibfield  {author} {\bibinfo {author} {\bibfnamefont {E.~H.}\ \bibnamefont
  {Fradkin}}\ and\ \bibinfo {author} {\bibfnamefont {S.~H.}\ \bibnamefont
  {Shenker}},\ }\href {\doibase 10.1103/PhysRevD.19.3682} {\bibfield  {journal}
  {\bibinfo  {journal} {Phys. Rev. D}\ }\textbf {\bibinfo {volume} {19}},\
  \bibinfo {pages} {3682} (\bibinfo {year} {1979})}\BibitemShut {NoStop}%
\bibitem [{\citenamefont {Banks}\ and\ \citenamefont
  {Rabinovici}(1979)}]{Banks:1979fi}%
  \BibitemOpen
  \bibfield  {author} {\bibinfo {author} {\bibfnamefont {T.}~\bibnamefont
  {Banks}}\ and\ \bibinfo {author} {\bibfnamefont {E.}~\bibnamefont
  {Rabinovici}},\ }\href {\doibase 10.1016/0550-3213(79)90064-6} {\bibfield
  {journal} {\bibinfo  {journal} {Nucl. Phys. B}\ }\textbf {\bibinfo {volume}
  {160}},\ \bibinfo {pages} {349} (\bibinfo {year} {1979})}\BibitemShut
  {NoStop}%
\bibitem [{\citenamefont {Seiler}(2015)}]{Seiler:2015rwa}%
  \BibitemOpen
  \bibfield  {author} {\bibinfo {author} {\bibfnamefont {E.}~\bibnamefont
  {Seiler}},\ }\href@noop {} {\  (\bibinfo {year} {2015})},\ \Eprint
  {http://arxiv.org/abs/1506.00862} {arXiv:1506.00862 [hep-lat]} \BibitemShut
  {NoStop}%
\bibitem [{\citenamefont {Lee}\ and\ \citenamefont
  {Zinn-Justin}(1972)}]{Lee:1972fj}%
  \BibitemOpen
  \bibfield  {author} {\bibinfo {author} {\bibfnamefont {B.~W.}\ \bibnamefont
  {Lee}}\ and\ \bibinfo {author} {\bibfnamefont {J.}~\bibnamefont
  {Zinn-Justin}},\ }\href {\doibase 10.1103/PhysRevD.5.3121} {\bibfield
  {journal} {\bibinfo  {journal} {Phys. Rev. D}\ }\textbf {\bibinfo {volume}
  {5}},\ \bibinfo {pages} {3121} (\bibinfo {year} {1972})}\BibitemShut
  {NoStop}%
\bibitem [{\citenamefont {Caudy}\ and\ \citenamefont
  {Greensite}(2008)}]{Caudy:2007sf}%
  \BibitemOpen
  \bibfield  {author} {\bibinfo {author} {\bibfnamefont {W.}~\bibnamefont
  {Caudy}}\ and\ \bibinfo {author} {\bibfnamefont {J.}~\bibnamefont
  {Greensite}},\ }\href {\doibase 10.1103/PhysRevD.78.025018} {\bibfield
  {journal} {\bibinfo  {journal} {Phys. Rev. D}\ }\textbf {\bibinfo {volume}
  {78}},\ \bibinfo {pages} {025018} (\bibinfo {year} {2008})},\ \Eprint
  {http://arxiv.org/abs/0712.0999} {arXiv:0712.0999 [hep-lat]} \BibitemShut
  {NoStop}%
\bibitem [{Note3()}]{Note3}%
  \BibitemOpen
  \bibinfo {note} {Strictly speaking, the vanishing of the Higgs condensate at
  $B>B_{c2}$ is not evident from the normalized condensate squared of Fig.~\ref
  {fig:PhiSQ}(a) without an additional analysis. Indeed, the quantity ${\left
  \langle \phi ^2 \right \rangle }$ is affected by significant ultraviolet
  contributions which are always present in broken and unbroken phases.
  However, the homogenisation of the $W$ condensate observed above $B_{c2}$
  suggests that the Higgs condensate should be absent at $B >
  B_{c2}$.}\BibitemShut {Stop}%
\bibitem [{\citenamefont {Ambjorn}\ and\ \citenamefont
  {Olesen}(1990{\natexlab{b}})}]{Ambjorn:1989sz}%
  \BibitemOpen
  \bibfield  {author} {\bibinfo {author} {\bibfnamefont {J.}~\bibnamefont
  {Ambjorn}}\ and\ \bibinfo {author} {\bibfnamefont {P.}~\bibnamefont
  {Olesen}},\ }\href {\doibase 10.1142/S0217751X90001914} {\bibfield  {journal}
  {\bibinfo  {journal} {Int. J. Mod. Phys. A}\ }\textbf {\bibinfo {volume}
  {5}},\ \bibinfo {pages} {4525} (\bibinfo {year}
  {1990}{\natexlab{b}})}\BibitemShut {NoStop}%
\bibitem [{\citenamefont {Ambjorn}\ and\ \citenamefont
  {Olesen}(1992)}]{Ambjorn:1992ca}%
  \BibitemOpen
  \bibfield  {author} {\bibinfo {author} {\bibfnamefont {J.}~\bibnamefont
  {Ambjorn}}\ and\ \bibinfo {author} {\bibfnamefont {P.}~\bibnamefont
  {Olesen}},\ }in\ \href@noop {} {\emph {\bibinfo {booktitle} {{4th Hellenic
  School on Elementary Particle Physics}}}}\ (\bibinfo {year} {1992})\ pp.\
  \bibinfo {pages} {396--406},\ \Eprint {http://arxiv.org/abs/hep-ph/9304220}
  {arXiv:hep-ph/9304220} \BibitemShut {NoStop}%
\bibitem [{Note4()}]{Note4}%
  \BibitemOpen
  \bibinfo {note} {We do not perform any smoothening or cooling procedures that
  could drive our Monte Carlo configurations of Figs.~\ref
  {fig:structure}(e)-(m) closer to the classical regime.}\BibitemShut {Stop}%
\bibitem [{\citenamefont {Ambjorn}\ and\ \citenamefont
  {Olesen}(1980)}]{Ambjorn:1979xi}%
  \BibitemOpen
  \bibfield  {author} {\bibinfo {author} {\bibfnamefont {J.}~\bibnamefont
  {Ambjorn}}\ and\ \bibinfo {author} {\bibfnamefont {P.}~\bibnamefont
  {Olesen}},\ }\href {\doibase 10.1016/0550-3213(80)90476-9} {\bibfield
  {journal} {\bibinfo  {journal} {Nucl. Phys. B}\ }\textbf {\bibinfo {volume}
  {170}},\ \bibinfo {pages} {60} (\bibinfo {year} {1980})}\BibitemShut
  {NoStop}%
\bibitem [{\citenamefont {Rasolt}\ and\ \citenamefont
  {Te\v{s}anovi\'c}(1992)}]{Rasolt1992}%
  \BibitemOpen
  \bibfield  {author} {\bibinfo {author} {\bibfnamefont {M.}~\bibnamefont
  {Rasolt}}\ and\ \bibinfo {author} {\bibfnamefont {Z.}~\bibnamefont
  {Te\v{s}anovi\'c}},\ }\href {\doibase 10.1103/RevModPhys.64.709} {\bibfield
  {journal} {\bibinfo  {journal} {Rev. Mod. Phys.}\ }\textbf {\bibinfo {volume}
  {64}},\ \bibinfo {pages} {709} (\bibinfo {year} {1992})}\BibitemShut
  {NoStop}%
\end{thebibliography}

%


\vskip 10mm

\centerline{{\bf\Large{Supplemental Material}}}

\vskip 3mm

\paragraph*{\bf Appendix A. Lattice Electroweak model in hypermagnetic field background.}  Most lattice simulations~\cite{Bunk:1992xt,Bunk:1992kf,Fodor:1994sj,Fodor:1994dm,Csikor:1996sp,Aoki:1999fi} of the electroweak model were performed at Higgs masses different from the physical value of the Higgs mass as these simulations were carried out before the Higgs particle was discovered experimentally. Moreover, since the main aim of most simulations was to find the location and the strength of the electroweak transition (the latter property is essential for baryogenesis~\cite{Sakharov:1967dj,Rubakov:1996vz}), a large number of simulations were done in a dimensionally reduced 3d lattice model valid at high temperatures~\cite{Kajantie:1993ag,Ilgenfritz:1995sh,Gurtler:1996wx,Kajantie:1996qd,Kajantie:1996mn,Kajantie:1998rz}. In addition, the hypercharge field $Y_\mu$ has been taken into account only in some studies~\cite{Kajantie:1996qd,DOnofrio:2015gop} because, in the absence of a background (hyper)magnetic field, the hypercharge sector plays a relatively modest role given the weakness of the hypercharge coupling $g'$ and the Abelian nature of this field. Since our aim is different from most of the previous works, we simulate the model (i) with the physical Higgs mass (ii) in the four-dimensional formulation (iii) with the hypergauge field included. All these factors are required to uncover the properties of the electroweak vacuum in the background magnetic field.

We use the following lattice action for the theory~\eq{eq:LEW}:
\beqn
S & = & \beta\us{x,\mu<\nu}{\sum}\lr{1 - \frac{1}{2}\Tr U_{x,\mu\nu}}
    + \frac{\beta_Y}{2}\us{x,\mu<\nu}{\sum}\, \theta_{x,\mu\nu}^2 \nonumber \\
  & + & \us{x}{\sum} \lr{-\kappa\phi_{x}^{\dagger}\phi_{x} + \lambda\lr{\phi_{x}^{\dagger}\phi_{x}}^2} 
\label{eq:S:lat}\\
  & + & \us{x,\mu}{\sum}\lrm{\phi_x - e^{i \lr{\theta_{x,\mu} + \theta_{x,\mu}^Y}} U_{x,\mu} \phi_{x+\hmu}}^2\,, \nonumber
\eeqn
where $\phi_{x}$ is the Higgs doublet of two complex scalar fields (an equivalent, widely used formulation of the Higgs field in terms of an SU(2) matrix~\cite{Langguth:1985dr} gives identical results). 

In the leading order of a small lattice spacing~$a$, the lattice SU(2) gauge field $U_{x,\mu}$ and the noncompact $U(1)$ hypercharge gauge field $\theta_{x,\mu}$ are related to their continuum counterparts $A^a_\mu$ and $Y_\mu$, respectively, as follows: $U_{x,\mu} = e^{i a t^a A^a_\mu(x)}$ and $\theta_{x,\mu} = a Y_\mu(x)/2$, where $t^a = \sigma^a/2$ are the generators of the SU(2) gauge group expressed via the Pauli matrices $\sigma^a$. The lattice plaquettes $U_{x,\mu\nu}=U_{x,\mu}U_{x+\hmu,\nu}U_{x+\hnu,\mu}^{\dagger}U_{x,\nu}^{\dagger}$ and $\theta_{x,\mu\nu}=\theta_{x,\mu}+\theta_{x+\hmu,\nu}-\theta_{x+\hnu,\mu}-\theta_{x,\nu}$ correspond to the field strengths~\eq{eq:W:munu} and \eq{eq:Y:munu}, respectively. Using a set of identifications, $\beta \to 4/g^2$ and $\beta_Y \to 4/{g'}^2$, as well as the rescalings $A_\mu \to g A_\mu$ and $Y_\mu \to g' Y_\mu$, we recover the continuum theory~\eq{eq:LEW} from the lattice action~\eq{eq:S:lat}.

The spatially uniform hypermagnetic field,
\beqn
{\bs B}_Y = (0,0,B_Y), \qquad B_Y = \frac{2}{g'} \cdot \frac{2 \pi k}{(L_s a)^2}, \qquad
\label{eq:H:lattice}
\eeqn
is expressed in a standard way. The integer $k \in \Z$ determines the total number $2 k$ of the elementary fluxes $\Phi_Y = 2 \pi/g'$ carried by the hypermagnetic field through a lattice cross-section. In the lattice action~\eq{eq:S:lat}, the hypermagnetic field~\eq{eq:H:lattice} is introduced via the shift of the lattice hypercharge field $\theta_{\text{x},\mu} \to \theta_{\text{x},\mu} + \theta_{\text{x},\mu}^Y$ similarly to the standard practices~\cite{Bali:2011qj}. The flux quantization~\eq{eq:H:lattice} appears as a result of periodic boundary conditions imposed in our calculations (more details are given in Appendix~B).

In addition to the external hypergauge field $\theta_{x,\mu}^Y$, we have four parameters of the theory: $\beta$, $\beta_Y$, $\kappa$ and $\lambda$. These parameters can be fixed via the requirement that the lattice model matches the known phenomenological parameters of the electroweak model at a zero magnetic field. As a matching criterion, we choose the mass ratios, $m_W/m_H$, and $m_Z/m_H$ (that also determine the Weinberg angle $\theta_W$), and one of the charges, $g$ or $g'$ (the other one is determined by the Weinberg angle $\theta_W$). A remaining degree of freedom corresponds to the variation in the lattice spacing $a$, which plays the role of an ultraviolet cutoff. The technical details highlighting our choice of lattice couplings are described in Appendix~C.

We employ the Hybrid Monte Carlo technique to generate lattice field configurations~\cite{Gattringer:2010zz}. For a zero value of the hypermagnetic field $B_Y$, we use about $16\times 10^6$ configurations in order to fix the physical scale. For each nonzero value of $B_Y$, we work with about $10^6$ configurations. We perform our calculations in the Euclidean lattice volume $L_t \times L_s^3$ with $L_t = 64$ and $L_s = 48$ corresponding to the zero-temperature limit. 

We also ensure that we work closely to the thermodynamic limit by performing calculations on different lattice volumes $L^4$ sizes with $L=16,\,24,\,32,\,48,\,64$ for all measured quantities in a selected set of points that mark transition and limiting regions in all phases. We checked that our results are independent of the lattice volume within statistical errors, thus signaling that our results are valid in the thermodynamic limit. 

\vskip 1mm
\paragraph*{\bf Appendix B: Hypermagnetic field.}  This article aims to study the behavior of the electroweak model~\eq{eq:LEW} in the background magnetic field ${\bs B}$. However, the very notion of the magnetic field ${\bs B} = {\bs \nabla} \times {\bs A}$ loses its physical meaning in a phase with the restored electroweak symmetry, which is suggested to be induced by this same field. Indeed, the definition of the electromagnetic gauge field $A^\mu = (A^0, {\bs A})$ (and, therefore, of $\bs B$) depends on the direction of the condensate of the Higgs doublet field $\avr{\phi}$ in the internal SU(2) space. For example, in the broken phase, the electromagnetic field~\eq{eq:A:field} is identified in the unitary gauge where the upper component of $\avr{\phi}$ vanishes while the lower component is kept nonzero. 
As we expect that the magnetic field can restore the electroweak symmetry, $\avr{\phi} \to 0$, the electromagnetic direction becomes an ill-defined quantity, thus questioning the very meaning of the magnetic field $\bs B$ itself. 

The flux quantization of an Abelian flux~\eq{eq:H:lattice} is the result of periodic boundary conditions imposed on the appropriate Abelian gauge field: the Wilson loop, which includes the whole cross-section of a periodically compactified space, equals a unity because the contributions from the opposite parallel segments of the loop cancel each other exactly. 

The loop is a product of the link transporters, which can be read off from the interaction of the gauge field with a matter field. In our case~\eq{eq:S:lat}, the link transporter is the phase $e^{i \theta_{x,\mu}}$, where, as we mentioned above, $\theta_{x,\mu} = a (g'/2) Y_\mu(x)$ in a continuum limit. The total flux~\eq{eq:H:lattice} in a periodic space can therefore introduce the hypermagnetic fluxes only in pairs (namely, $2 k$ vortices for $k \in \mathbb{N}$) due to the presence of the $g'/2$ factor in the interaction term~\eq{eq:D:mu}. A similar situation also appears in QCD, where the fractional electric charge of quarks, $+ 2e/3$ and $-2e/3$, leads to the quantization of the magnetic fluxes in triples~\cite{Bali:2011qj}.

Thus, the flux quantization of the hypermagnetic flux (and, therefore, the value of a possible minimal nonzero flux) is a result of the periodic boundary conditions rather than a consequence of the lattice discretization. The latter introduces an ultraviolet cutoff which sets an upper bound, $k_{\mathrm{max}} = L^2/2$, on the total number $k$ of {\it pairs} of the elementary hypermagnetic fluxes that can be introduced on the lattice: $k = 0, 1, \dots , L^2/2$.

We introduce the hypermagnetic field~\eq{eq:H:lattice} as a hypercharge field $\theta_{\text{x},\mu}^Y$ similarly to the standard practices~\cite{Bali:2011qj}:
\beqn
  \theta_{x,2}^Y(k) = \frac{2\pi k}{L^2_s} x, \quad
  \theta_{x,1}^Y(k)|_{x = L_s - 1} = -\frac{2\pi k}{L_s} y, \qquad\
  \label{eq:theta:B}
\eeqn
with $\theta_{x,3}^Y = \theta_{x,4}^Y = 0$. Alternatively, one could introduce the hypermagnetic field via a ``chemical potential'' for the magnetic flux~\cite{Kajantie:1998rz}.

Notice that the emergence of possible mixed phases is not disfavored by the uniform nature of the hypermagnetic background field~\eq{eq:H:lattice} since the background field fixes only the global flux through the whole lattice cross-section without affecting the local structures corresponding to possible inhomogeneities in the field.

\vskip 1mm
\paragraph*{\bf Appendix C: Masses and physical point.}  We fix the ratio of the masses of the Higgs particle as well as the $Z$ and $W$ bosons to their physical values at a vanishing hypermagnetic field. To this end, we calculate the following slice-slice correlation functions:
\beqn
\label{eq:corH}
C^{H}(l) & = & \frac{1}{L_t} \sum_{t=0}^{L_t-1} \vev{\rho^2(t)\cdot \rho^2(t+l)}\,,\\
\label{eq:corZ}
C^{Z}_{\mu}(l) & = & \frac{1}{L_t} \sum_{t=0}^{L_t-1}  \vev{z_{\mu}(t)\cdot z_{\mu}(t+l)}\,, \\
\label{eq:corW}
C^{W}_{\mu}(l) & = & \frac{1}{L_t} \sum_{t=0}^{L_t-1}  \vev{w^{*}_{\mu}(t)\cdot w_{\mu}(t+l)}\,,
\eeqn
where the gauge-invariant quantities, 
\beqn
\rho^2(t) & = & \frac{1}{V_s}\us{x\in V_s}{\sum} \lr{\phi^{\dagger}_x\phi_x}(t)\,,
\label{eq:phi}\\
z_{\mu}(t) & = & \frac{1}{V_s}\us{x\in V_s}{\sum} z_{x,\mu}(t)\,,
\label{eq:z}\\
w_{\mu}(t) & = & \frac{1}{V_s}\us{x\in V_s}{\sum} U^{12}_{x,\mu}(t)\,,
\label{eq:w}\\
z_{x,\mu} & = & \Arg\lr{\phi^{\dagger}_x e^{i \lr{\theta_{x,\mu}+\theta_{x,\mu}^Y}} U_{x,\mu} \phi_{x+\hmu}}\,,
\label{eq:zxmu}
\qquad
\eeqn
are given by the spatial averaging over the spatial volume $V_s = L_s^3$ for each time slice fixed by the Euclidean time~$t$.

The use of the operators~\eq{eq:phi}-\eq{eq:zxmu} requires a word of caution, because these definitions of physical fields are not, strictly speaking, gauge-invariant quantities which are used in the perturbation theory to determine corresponding particle fields. The use these simplified operators is justified by the fact that the discrepancy between the definitions is irrelevant in the leading order of perturbation theory~\cite{Frohlich:1980gj,Frohlich:1981yi} (see also Refs.~\cite{Woloshyn:2017rhe,Lewis:2018srt} for the photon operators with the proper discrete quantum numbers). The reader can find the excellent review on the subject in Ref.~\cite{Maas:2017wzi}.

We extract the physical Higgs mass $m_H$ from the correlation function~\eq{eq:corH} of the $\rho^2$ field~\eq{eq:phi} following Ref.~\cite{Langguth:1985dr}. The definition of the lattice $Z$-boson  field~\eq{eq:z} in the correlator~\eq{eq:corZ} comes from the form of the interaction term of the lattice action~\eq{eq:S:lat}. It corresponds to the continuum neutral vector $Z$ boson  field~\eq{eq:Z:field}, which is rigidly fixed by the covariant derivative~\eq{eq:D:mu}, which appears in the continuum action~\eq{eq:LEW}. The two-point correlation function~\eq{eq:corW} includes the off-diagonal component $U^{12}_{x,\mu}$ of the $SU(2)$ matrix $U_{x,\mu}$ which corresponds to the physical $W$ field in the unitary gauge. In our calculations, we fix the unitary gauge for the $SU(2)_W$ subgroup supplemented by the maximal tree gauge for the $U(1)_Y$ subgroup. The gauge fixing is needed to correctly identify the $W$-boson field and reduce the noise associated with the residual Abelian gauge freedom. 

To reduce the effect of the perturbative noise further, we use the spatial APE-smearing procedure~\cite{Durr:2004xu} with parameters $\alpha_{\mathrm{APE}}=0.5$ and $n_{\mathrm{APE}}=100$ with respect to the $U_{x,\mu}$ and $\theta_{x,\mu}$ fields.

The sequence of the actions in the calculation of the $W$-boson two-point function is as follows. First, we perform the spatial APE smearing for SU(2) and U(1) gauge fields, then we use a maximal tree gauge fixing for the U(1) gauge field, and then we utilize the unitary gauge for the SU(2) gauge field. Finally, we calculate the two-point function.

\begin{figure}[!htb]
\begin{center}
\includegraphics[width=0.45\textwidth,clip=true]{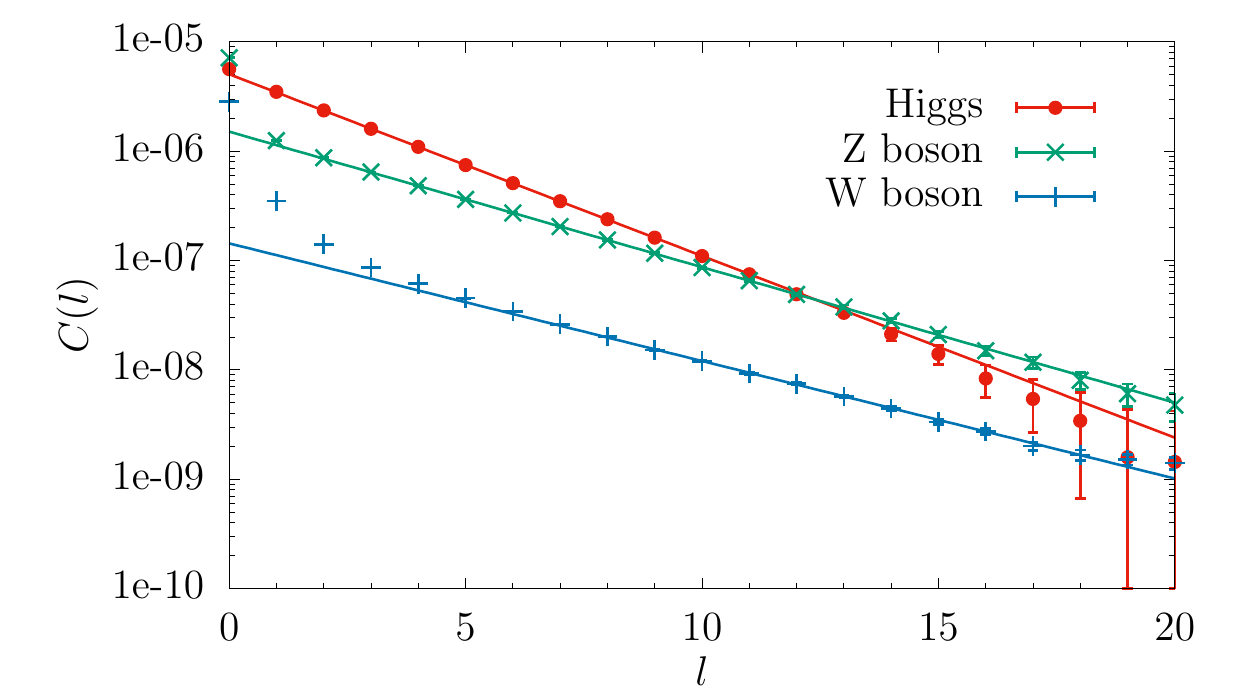}
\end{center}
\vskip -4mm 
\caption{The correlation functions for the Higgs~\eq{eq:corH}, $Z$-boson ~\eq{eq:corZ}, and $W$-boson  (\ref{eq:corW}) fields that are used for calculating their masses at a vanishing background hypermagnetic field, $B_Y = 0$, via the fits by the function~\eq{eq:fit:corW}}.
\label{fig:CorrelHZW}
\end{figure}

The numerical results for the two-point functions~\eq{eq:corH}, \eq{eq:corZ} and \eq{eq:corW} are then fitted by the exponential function,
\beqn
\label{eq:fit:corW}
C(l)=C_0\lr{e^{-M\, l}+e^{-M\, (L_t - l)}} + C_1\,,
\eeqn
where the amplitude $C_0$, the (squared) condensate $C_1$, and the lattice mass $M$ are the fit parameters. The lattice mass $M \equiv M^{\mathrm{lat}} = M^{\mathrm{phys}} a$ is related to the physical mass $M^{\mathrm{phys}}$ via the lattice spacing $a$. The examples of the fits~\eq{eq:fit:corW} of all three correlators \eq{eq:corH}, \eq{eq:corZ} and \eq{eq:corW} are provided in Fig.~\ref{fig:CorrelHZW}. Table~\ref{tab:params} shows the masses and corresponding to the set of parameters used in the paper.

The correct choice of the ratios of the W, $Z$, and Higgs boson masses is essential for determining the correct structure of the vortex ground state~\cite{Skalozub:1986gw,Ambjorn:1989bd,MacDowell:1991fw}. 

The determination of the SU(2) gauge coupling seems to provide us with the uncertainty that has been, fortunately, resolved a few decades ago. In the literature, one customarily sets $\beta = 8$ for the SU(2) lattice coupling since this value corresponds to the renormalized $SU(2)$ gauge coupling $g^2_R \simeq 0.5$, which, in turn, lies closely to its phenomenological value in continuum limit~\cite{Langguth:1985dr,Bunk:1992xt,Fodor:1994dm}. We also explored other values of the SU(2) coupling constant. We ensured that all the conclusions of our paper are independent of the actual value of $\beta$ in a wide set of values, provided the mentioned mass ratios are kept fixed. This property is valid even on the quantitative level, including the positions of the phase transitions. In our paper, we use $\beta=12$, for which we have the best statistics. In the paper, we use the set of the lattice parameters shown in Table~\ref{tab:params}. 

\begin{table}[]
    \centering
    \begin{tabular}{|c|c|c|c|c|c|c|}
    \hline 
$\beta$ & $\beta_Y$ & $\lambda$ & $\kappa$ & $m_H a$ & $m_Z,\GeV$ & $m_W,\GeV$ \\ 
    \hline 
$12$ & $10.45$ & $0.175$ & $0.4030$ & $0.3827(9)$ & $93.42(24)$ & $81.1(6)$ \\ 
    \hline 
\end{tabular} 
\caption{Parameters of the lattice Lagrangian~\eq{eq:S:lat}, the corresponding lattice spacing $a$ (expressed in terms of the Higgs mass $m_H=125.25(17)\GeV$~\cite{PDG}), and the masses of the $W$ and $Z$ bosons. Set 1 corresponds to the physical point (given to the Particle Data Group~\cite{PDG}, $m_W = 80.377(12)\GeV$, $m_Z = 91.1876(21)\GeV$). 
The Weinberg angle~$\theta_W$ equals to its physical value with the relative error of less than $2\%$.}
    \label{tab:params}
\end{table}

The value of the lattice spacing for the physical point allows us to achieve the resolution of the hypermagnetic field about $10\%$ in units of the first critical field $B_{c1}$. The finite resolution, which appears as a result of the quantization of magnetic flux, provides us with the systematic uncertainty in estimating the values of the critical magnetic fields~\eq{eq:Bc1:lat} and \eq{eq:Bc2:lat}. 

In our simulations, we choose the lattice Higgs mass around the value $m_H a\simeq 0.38$, which corresponds to the lattice $W$-boson  mass $m_W a \simeq 0.25$ at a vanishing background field $B_Y = 0$. For our lattice $64\times 48^3$, this choice of the lattice spacing allows us to reach sufficient accuracy for all mass correlators while, at the same time, making it possible to achieve an acceptably small value of elementary hypermagnetic flux that determines the distance between the nearest values of the hypermagnetic field~\eq{eq:H:lattice}. The latter factor is crucial for the precise determination of the magnetic-field-induced transitions with sufficient accuracy because of the elementary flux given the computational restrictions imposed on the maximal lattice volume. 

The small value of lattice spacing also ensures the closeness of the continuum limit where the rotational and translational symmetries, broken by the lattice, get restored. Such limits were studied in various approaches (we mention here the earlier studies of the electroweak endpoint~\cite{Csikor:1998eu} and recent continuum extrapolation of the electroweak crossover~\cite{DOnofrio:2015gop}). One should also notice that, similarly to Ref.~\cite{DOnofrio:2015gop}, we are working far from the Landau pole so that the hypergauge $U(1)$ part is kept in the perturbative regime and the limit of diminishing lattice spacing, $a \to 0$, can be properly addressed.

The values of our quartic coupling $\lambda$ and the parameter $\kappa$, shown in Table~\ref{tab:params}, are different from typical values used in the literature to investigate the finite-temperature phase transition in the dimensionally unreduced, 3+1 dimensional model. For example, after an adaptation to our notations, the quartic coupling used in a finite-temperature study of Ref.~\cite{Csikor:1996sp} lies in the $\lambda \simeq 0.02$, which is an order of magnitude smaller than our value. This difference can partially be explained by the fact that we work with a heavier (in physical units) Higgs particle. In addition, we aim to decrease the gap  $\delta B_Y = 4 \pi/(g' L^2 a^2)$ between the nearest values of the hypermagnetic field strengths~\eq{eq:H:lattice} in order to resolve better the evolution of the vacuum properties with increasing hypermagnetic field. To this end, we need to increase the physical size of the spatial lattice, which implies, in particular, augmenting the physical length of the lattice spacing. On the other hand, we cannot set $a$ too large since it would lead to substantial ultraviolet artifacts. In finite-temperature studies, one tends, on the contrary, to choose the lattice spacing as small as possible in order to increase temperature and to reach the symmetry-restored part of the wide electroweak crossover. 

Our choice of lattice spacing provides a compromise between these two purely technical constraints. Moreover, the slight deviation of the $m_Z$ and $m_W$ masses from their phenomenological values is already of the order of the systematic resolution of the hypermagnetic field in our work. Therefore, we conclude that the present accuracy is sufficient for our purposes related to determining the approximate values of the critical fields and the nature of the new, magnetic-field-induced phases in the model.

\vskip 1mm
\paragraph*{\bf Appendix D: Lattice observables.}  The evolution of the expectation value of the Higgs field squared with the magnetic field, Fig.~\ref{fig:PhiSQ}(a), as well as the local excess of the Higgs expectation value in the transverse $(x,y)$ plane, Fig.~\ref{fig:structure}(i), have been numerically calculated using the following gauge-invariant operator:
\begin{eqnarray}
  \phi^2({\bs x}) & = & \frac{1}{N_t N_z}\underset{t,z}{\sum} \phi_x^\dagger\phi_x\,.
  \label{eq:phi2:appendix}
\end{eqnarray}
We denote ${\bs x} = (x,y)$ the coordinate in the longitudinal plane. The sum in the longitudinal $(t,z)$ plane is applied to all observables to diminish statistical fluctuations and increase the signal-to-noise ratio. This averaging is applied to all numerically calculated quantities in Fig.~\ref{fig:structure}. To visualize the local Higgs structure in Fig.~\ref{fig:structure}(i), we took a square of the local value of the condensate~\eq{eq:phi2:appendix}.

The (squared) components of the $W$ field, $W^2_\parallel$ and $W^2_\perp$, are plotted as a function of the background magnetic field in Fig.~\ref{fig:W2}. The spatial behavior of the same components, $W_\parallel$ and $W_\perp$, is visualized in Figs.~\ref{fig:structure}(e) and \ref{fig:structure}(g), respectively. We calculate the squared condensates with the help of the gauge-invariant operator,
\beqn
  & & W_{\mu}^{2}({\bs x}) = \frac{1}{N_t N_z}\underset{t,z}{\sum} w_{x,\mu}\,,\\
  & & w_{x,\mu} = \frac{1}{2} \bigl[(1 - \frac{1}{2}\Tr\lr{{\hat n}_x U_{x,\mu} {\hat n}_{x + {\hat \mu}} U_{x,\mu}^{\dagger}}) \bigr], \qquad
  \label{eq:W:induced:gauge}
\eeqn
where we introduced the auxiliary matrix field: 
\beqn
{\hat n}_x & = & n_x^a \sigma^a \equiv {\vec n} \cdot {\vec \sigma}\\
n_x^a & = & - \frac{\phi_x^\dagger \sigma^a \phi_x}{\phi_x^\dagger \phi_x}.
\eeqn
The field~\eq{eq:W:induced:gauge} is invariant with respect to the $U(1)_Y$ hypergauge subgroup of the entire electroweak group. Under the subgroup of SU(2) gauge transformations, the fields transform as follows:
\beqn
SU(2)_W: \qquad 
\left\{
\begin{array}{rcl}
U_{x,\mu} & \to & \Omega_x U_{x,\mu} \Omega^\dagger_{x+ {\hat \mu}}, \\
\phi_x & \to & \Omega_x \phi_x, \\
{\hat n}_x & \to & \Omega_x {\hat n}_x \Omega^\dagger_x,
\end{array}
\right. \qquad
\label{eq:SU2:transformations}
\eeqn
implying that the field~\eq{eq:W:induced:gauge} is a gauge-invariant quantity.

The meaning of the construction~\eq{eq:W:induced:gauge} becomes clear in the unitary gauge, $\phi_x = (0,\rho_x)^T$, where the adjoint vector ${\hat n}_x$ points out to the third direction, ${\hat n}_x = \sigma_3$. In this gauge, the off-diagonal components of the $W$ boson field $W_\mu = W^a_\mu \sigma^a/2$ correspond to the standard charged $W$ bosons: $W^\pm_\mu = (W^1_\mu \mp i W^2_\mu)/\sqrt{2}$. Representing the SU(2) gauge field $U_{x,\mu}$ as an exponent of the continuum $W$ field,  $U_{x,\mu} = \exp\{i \frac{1}{2} a W_\mu^a \sigma^a\}$, and expanding Eq.~\eq{eq:W:induced:gauge} in terms of the lattice spacing $a$, one gets the $W$ condensate squared, $|W_\mu|^2 = W^+_{x,\mu} W^-_{x,\mu}$:
\beqn
w_{x,\mu} = \frac{1}{2} a^2 W^+_{x,\mu} W^-_{x,\mu} + O(a^4)\,.
\label{eq:w:W2}
\eeqn
Here, no sum over the repeating index $\mu$ is implemented. Thus, in the continuum limit, the composite field~\eq{eq:W:induced:gauge} is proportional to the $W$ condensate (squared)~\eq{eq:w:W2}. While the unitary gauge has been assumed in the derivation of Eq.~\eq{eq:w:W2}, the expression~\eq{eq:W:induced:gauge} is a gauge-invariant quantity implying that the association~\eq{eq:w:W2} of $w_{x,\mu}$ with the $|W_\mu|^2$ condensate works in any gauge.

In Fig.~\ref{fig:W2}, we give the $W$ condensate in the leading order in the lattice spacing $a$:
\beqn
W_\perp^2 \equiv \frac{1}{a^2} \left( w_{x,1} + w_{x,2} \right)\,.
\eeqn

The $Z$ flux in Fig.~\ref{fig:PhiSQ}(f) is defined, similarly to the $W$ condensate, as a gauge-invariant expression:
\beqn
  Z_{12}({\bs x}) & = & \frac{1}{N_t N_z}\underset{t,z}{\sum} Z_{x,12}\,,\\
  Z_{x,\mu\nu} & = & \varphi_{x,\mu}+\varphi_{x+\hmu,\nu}-\varphi_{x+\hnu,\mu}-\varphi_{x,\nu}\,,\qquad \\
  \varphi_{x,\mu} & = & \Arg\lr{e^{i (\theta_{x,\mu}+\theta_{x,\mu}^Y)} \phi_{x}^{\dagger} U_{x,\mu} \phi_{x+\hmu}}\,. \qquad
\label{eq:varphi}
\eeqn
In the continuum limit, Eq.~\eq{eq:varphi} reduces, up to a charge-related factor, to the usual expression of the $Z$ flux expressed via the $Z$ boson field~\eq{eq:Z:field}: $Z_{12} = \partial_1 Z_2 - \partial_2 Z_1$.To see this fact, we notice that the scalar Higgs field $\phi$ does not obviously carry electric charge (so that it does not couple to the electromagnetic field $A_\mu$) and it couples only to the neutral Z boson~\eq{eq:Z:field}. In the continuum formulation of the Electroweak theory, this property is realized via the fact that the lower component of the scalar field $\phi$ in the unitary gauge interacts, via the kinetic term, only with the $Z_\mu$ vector~\eq{eq:D:mu}. In consistency with the continuum definition, the lattice vector field which interacts with the lower component of the lattice scalar field in the unitary gauge, corresponds to the lattice version of the Z boson field. Substituting $\phi_x = (0,\rho_x)^T$ into the kinetic term of the scalar field in the lattice action~\eq{eq:S:lat}, we recover that the interaction part of the kinetic term contains, apart from the trivial $\rho_x^2 + \rho_{x+\hat\mu}^2$ contribution, the interaction with the vector field: $\rho_{x} e^{i\varphi_{x,\mu}} \rho_{x+\hat\mu} + c.c.$, where the lattice version of the Z boson field coincides exactly with our definition~\eq{eq:varphi}. Similarly to our construction of the $W$ operator~\eq{eq:W:induced:gauge}, the expression for the Z field~\eq{eq:varphi} is a gauge-invariant quantity implying that this lattice construction is valid in any gauge.

The fields $Z_\parallel$ and $Z_\perp$ in Fig.~\ref{fig:PhiSQ}(h) and (l), respectively, are the longitudinal and the transverse components of the $Z$ field defined via the lattice angle $\varphi$, Eq.~\eq{eq:varphi}:
\begin{eqnarray}
  Z_{\mu}({\bs x}) & = & \frac{1}{N_t N_z}\underset{t,z}{\sum} \varphi_{x,\mu}\,.
\end{eqnarray}

The excess of the electromagnetic field,
\begin{eqnarray}
  \Delta B({\bs x}) & = & \frac{1}{N_t N_z}\underset{t,z}{\sum} {\bar \theta}^{\EM}_{x,12}\,,
\end{eqnarray}
shown in Fig.~\ref{fig:structure}(k), is expressed via the electromagnetic gauge-invariant flux:
\beqn
  {\bar \theta}^{\EM}_{x,\mu\nu} & = & \arg \Bigl( e^{i  \theta_{x,\mu\nu}} \tr
  \Bigl[\frac{1}{2}(\bbbone + n_x) V_{x,\mu\nu} \Bigr]\Bigr)\,, \qquad
  \label{eq:AP:2}\\
  \theta_{x,\mu\nu} & = & \theta_{x,\mu} + \theta_{x +\hat\mu,\nu} - \theta_{x + \hat\nu,\mu} - \theta_{x,\nu}\,,\qquad\\
  V_{x,\mu\nu} & = & V_{x,\mu} V_{x +\hat\mu,\nu} V^\dagger_{x + \hat\nu,\mu} V^\dagger_{x,\nu}\,,\\
  V_{x,\mu} & = & \frac{1}{2} \bigl(U_{x,\mu} + n_x U_{x,\mu} n_{x + \hat \mu}\bigr)\,.
  \label{V}
\eeqn
In the unitary gauge, the flux~\eq{eq:AP:2} is reduced to the standard expression of the magnetic field in terms of the photon gauge field~\eq{eq:A:field}. Being a gauge-invariant expression, Eq.~\eq{eq:AP:2} gives us the magnetic flux without a need for gauge fixing.

Finally, the current ${\bs J}_{\perp}^Z$ in Fig.~\ref{fig:structure}(m) is identified as a weak current coupled to the hypercharge field:
\beqn
  J_{\mu}^{(\theta)}({\bs x}) & = & - \frac{2}{N_t N_z}  \\
  & & \times \underset{t,z}{\sum}  {\mathrm{Im}}\lr{e^{i (\theta_{x,\mu}+\theta_{x,\mu}^Y)} \phi_{x}^{\dagger} U_{x,\mu} \phi_{x+\hmu}}. \qquad \nonumber
\eeqn
This neutral current corresponds to the variation of the matter action in the lattice electroweak Lagrangian~\eq{eq:S:lat} with respect to the lattice hypergauge vector field~$\theta$.

It is important to stress that the choice of gauge is crucial to determine the phases and the spectra. In many models that experience the effect of spontaneous symmetry breaking, the broken and unbroken phases are analytically connected by a path that circumvents an endpoint via a smooth crossover. Therefore, these two ``phases'' are, strictly speaking, parts of the same phase, implying the absence of the proper order parameter in a strict mathematical sense. This problem is pertinent to theories with a complex scalar field, such as the Electroweak model, and can be rigorously addressed in the lattice regularization of gauge theories~\cite{Osterwalder:1977pc,Fradkin:1978dv,Banks:1979fi,Seiler:2015rwa}. Moreover, as it was noted in the series of papers by Lee and Zinn-Justin~\cite{Lee:1972fj}, the detection of the phase transition depends on the operator and the gauge used, and the result can even differ between different gauges. Moreover, there are even gauges where the signatures of the transition are absent at all~\cite{Caudy:2007sf}.

\vskip 1mm
\paragraph*{\bf Appendix E: Illustration video.} \quad

\noindent
\href{https://www.youtube.com/watch?v=-TCZcLZWIBk}{A short video overview of the results: \\
https://www.youtube.com/watch?v=-TCZcLZWIBk.}

\end{document}